\documentclass[12pt,tightenlines,eqsecnum,floats,aps,amsmath,amssymb,nofootinbib,prd,superscriptaddress]{revtex4}

\usepackage{setspace}
\usepackage{amsmath,amssymb,amsfonts,amsthm,mathrsfs}
\usepackage{graphicx}
\usepackage{enumerate}

\def\be{\begin{equation}}
\def\ee{\end{equation}}
\def\ba{\begin{eqnarray}}
\def\ea{\end{eqnarray}}
\def\bi{\begin{itemize}}
\def\ei{\end{itemize}}

\def\bra{\langle}
\def\ket{\rangle}

\begin{document}

\title{Propagation in
Polymer Parameterised Field Theory}
\author{Madhavan Varadarajan}\email{madhavan@rri.res.in}
 \affiliation{Raman Research Institute, Bangalore 560080, India}

\begin{abstract}

The Hamiltonian constraint operator in Loop Quantum Gravity acts ultralocally.
Smolin has argued that this ultralocality seems incompatible with the existence
of a quantum dynamics which propagates perturbations between macroscopically seperated regions
of quantum geometry. We present evidence to the contrary within
an LQG type `polymer' quantization of two dimensional  Parameterised Field Theory (PFT). 
PFT is a generally covariant reformulation of free field propagation 
on flat spacetime. We show explicitly that while, as in LQG, the Hamiltonian 
constraint operator in PFT acts  ultralocally, states in the joint kernel of the Hamiltonian and 
diffeomorphism constraints of PFT {\em necessarily} describe propagation effects.
The particular structure of the finite triangulation 
Hamiltonian constraint operator plays a crucial role, as does the necessity of 
imposing (the continuum limit of) its kinematic adjoint as a constraint. Propagation is seen
as a property encoded by {\em physical} states in the kernel of the constraints rather than that of
repeated actions of the finite triangulation Hamiltonian constraint on kinematic states.
The analysis yields robust structural lessons for 
putative constructions of the Hamiltonian constraint in LQG for which ultralocal action
co-exists with a 
description of  propagation effects by physical states.

\end{abstract}
\maketitle

\section{Introduction} \label{sec1}

A key open problem in Loop Quantum Gravity is a satisfactory definition of the Hamiltonian 
constraint operator which generates its quantum dynamics. A mathematically well defined procedure to construct
this operator using canonical quantization techniques was developed by Thiemann \cite{tt}
\footnote{This work itself was based on cumulative progress achieved by many workers; for a detailed bibliography
see, for e.g. the review article \cite{lqgreview} and the book \cite{lqgbooks}}.
While signifiant progress has been made since Thiemann's pioneering work
the construction still yields an operator which is far from unique. Therefore it is desireable to subject
candidate operators to further requirements so as to cut down on the quantization choices responsible
for this non-uniqueness. For example, in \cite{hannojurek} a Hilbert space is constructed which supports the 
action of the operator so that the operator may be confronted with adjointness properties. Another example
is of recent work  which attempts to impose the requirement that the candidate 
operator yield an anomaly free algebra of constraints so as to impose spacetime covariance in the quantum 
theory \cite{aloku1,ctme,alok}. 

Here we are interested in yet another requirement, namely that the 
quantum dynamics describe the propagation of perturbations from one part of the quantum geometry to another.
This requirement was articulated several years ago by Smolin \cite{lee}. In his work, Smolin also 
offers a critique of Thiemann's general construction and concludes that such a construction method yields 
a constraint action which is 
`too local' to allow for propagation effects in the quantum theory. More in detail, 
recall that quantum states in LQG are superpositions of `spin network' states labelled by graphs. 
Each such state describes 
1d excitations of spatial geometry along the edges of the  graph which underlies the state. 
All known versions of the Hamiltonian constraint which derive from the general Thiemann procedure  act 
only at  vertices of the graph and the action
at one vertex is independent of the action at neighbouring vertices. Further, repeated actions of the Hamiltonian
constraint create a finer and finer nested structure about each vertex. Thus, modifications of the graph
structure wrought by the action of the Hamiltonian constraint in the immediate vicinity of one vertex do not 
seem to propagate to other vertices. Since such propagation is thought of as the non-perturbative
seed for  graviton propagation in semiclassical states in LQG, one would like the quantum
dynamics to allow for such propagation. 
Since the issue of the (semi)classical
limit of LQG is still open, Smolin's criticisms are  based, at least partly, on physical intuition. 
Nevertheless, these criticisms seem compelling and this article seeks to address them.

Given the complications of full blown gravity and the open issues within LQG, we focus on Smolin's criticisms
in the context of a generally covariant field theoretic toy model which has already proven to be extremely useful
in addressing other issues concerned with the Hamiltonian constraint. The model is that of 1+1 Parameterised Field Theory
(PFT) which is a generally covariant reformulation of free massless scalar field theory on 1+1 Minkwoski spacetime
in which, in addition to the scalar field, 
the Minkwoskian coordinates are treated as dynamical variables to be varied in the action.
Since on the one hand the classical theory describes scalar wave {\em propagation}, and on the other,
a complete LQG type `polymer' quantization exists for the model \cite{polypft1,polypft2,polypft3}
this model serves as an ideal testing ground
for an analyis and possible resolution of Smolin's criticisms. By virtue of its general covariance, the dynamics
of PFT is driven by Hamiltonian and  diffeomorphism constraints. The kinematic Hilbert space is spanned by
`charge network states'. Each such state is  labelled by a 1d graph on the Cauchy slice, the 
edges of the graph being colored by integer value charges  associated with 
quantum excitations of the Minkowsian coordinates and the  scalar field.
The physical Hilbert space can be constructed by group averaging techniques \cite{grpavge} and, despite 
being a rigorous quantization of {\em continuum} scalar field
theory, these physical states describe quantum scalar field excitations on {\em discrete} spacetime.
More in detail, a superselection sector of such states exists which describes quantum scalar field excitations {\em propagating} on a discrete spacetime 
{\em lattice}, the lattice spacing being governed by the analog of the Barbero- Immirzi parameter \cite{fergiorgio}.
For reasons explained in \cite{polypft2} we call this sector the `finest lattice' sector. 
The Hamiltonian constraint operator can also be constructed following the broad procedure introduced by Thiemann and 
quantization choices can be made  in such a way that its action is an infinitesmal version of the finite
transformations used in the group averaging procedure referred to above  (see \cite{polypft3} and Section \ref{sec4}).
The Hamiltonian constraint operator so constructed acts ultralocally.  By this we mean that, as in LQG, 
the  finite triangulation Hamiltonian constraint
\footnote{\label{f2}Recall that the   
Thiemann procedure introduces a set of finer and finer triangulations of the Cauchy slice, 
chooses at each triangulation an approximant to  the constraint constructed from the basic holonomy-flux functions of the 
theory so that at infinitely fine triangulation the approximant becomes exact, replaces the classical functions by 
quantum operators, and, finally, defines the resulting operator in a limit of  infinitely fine triangulation.}
acts on vertices of kinematical states, its action at a vertex 
is only sensitive to structure in a small neighbourhood of that vertex
and its action at one 
vertex is independent of that at other vertices.
As a result, its repeated action at a vertex does not lead by itself to
any propagation for reasons similar to that articulated by Smolin in the context of LQG.

Nevertheless, as we  show in this work, despite this ultralocal action, 
the joint kernel of the  Hamiltonian constraint operator
and the diffeomorphism constraint operator necessarily contains the finest lattice sector of physical states which 
{\em do} provide a description of propagation effects. Thus our work shows that propagation is {\em not} to be seen
as a property of repeated actions of the (finite triangulation) Hamiltonian constraint but rather as a (logically
independent) property
encoded in physical states which are in the joint kernel of the Hamiltonian and diffeomorphism constraints.
With this shift in focus to the structure of physical states in the kernel of the constraints, 
the two key features which enable propagation effects turn out to be\\
\noindent (i) the structure  of the Hamiltonian constraint at finite triangulation which is quite different from
the structure of Thiemann's Hamiltonian constraint despite their shared property of ultralocality.\\
\noindent (ii) the imposition of the continuum limit of the finite triangulation Hamiltonian constraint as well
as the continuum limit of the adjoint of this finite triangulation constraint, as operator constraints for physical 
states. 

We believe that the emphasis on propagation as a property encoded in physical states together with 
the general structural lessons learnt from PFT are robust and applicable to LQG and 
offer a way out of Smolin's criticisms. In other words, while Smolin's criticisms seem to hold for Thiemann's
choice of Hamiltonian constraint and while this choice leads to an ultralocal action, it does not follow
that the same criticisms need hold for other choices even if these choices also lead to an ultralocal action.
The obstacle to propagation is then {\em not} ultralocality. Hence, while the Thiemann procedure does seem to lead
to ultralocal constraint action, we are optimistic that there exist choices of
constraints constructed via the general Thiemann procedure which co-exist with physical states describing 
propagation effects.
\footnote{\label{fn3}
Preliminary work \cite{me} on the `$U(1)^3$' model \cite{leeg=0,alok,ctme} indicates that the notion of
physical states should be further restricted by demanding that they lie not only in the kernel of the Hamiltonian
and diffeomorphism constraints but also in that (of certain combinations of the Hamiltonian and) the `electric' 
diffeomorphism constraints, the latter being obtained by smearing the diffemorphism constraint by triad 
dependent shifts \cite{ctme}. It is not clear to us if this further restriction in the context of Thiemann's
specific choices in \cite{tt} suffices to yield physical propagation effects or if a different choice
of the Hamiltonian constraint is required in addition to this restriction; at the moment we believe the latter 
is more likely.}

The layout of this paper is as follows. In section \ref{sec2} we provide a brief review of classical PFT and its
polymer quantization wherein physical states are constructed through group averaging techniques.
The interested reader may consult \cite{polypft1,polypft2} for details of the formalism.
In section \ref{sec3} we focus on a particular physically relevant superselection sector of the Hilbert space 
known as the `finest lattice sector'. We 
 review the pictorial interpretation of kinematic and group averaged physical states in this sector in 
terms of discrete slices
on a spacetime lattice  carrying quantum matter excitations.   In section \ref{sec4} we review the action of the Hamiltonian
constraint from \cite{polypft3} and show that its action is ultralocal and that its repeated action on a finest lattice
charge net  does not 
cause long range propagation exactly as anticipated by Smolin. In the pictorial interpretation of section \ref{sec3}
this repeated action  fails to evolve the discrete Cauchy slice and its data
beyond a single lattice step. 
It turns out that the key to long range propagation is getting beyond one lattice step to two lattice steps.
In section \ref{sec5} we show how elements in  the joint kernel of the Hamiltonian and diffeomorphism constraints
encode evolution beyond a single lattice step to a second lattice step. 
We are able to isolate the key structural properties ( see (i) and (ii) above) of the Hamiltonian
constraint responsible for this.  This is the main result of this paper.
Once we have demonstrated evolution beyond a single lattice step to two lattice steps, a technical proof
can be constructed to show that long range evolution is encoded in physical states. We relegate this proof to the 
appendix as it relies on certain detailed technicalities discussed in \cite{polypft1,polypft2} for which familiarity
is assumed. Section \ref{sdiscuss} contains a discussion of our results in the context of LQG.

\section{Review of Polymer PFT \label{sec2}}
\subsection{Classical Theory\label{sct}}

The action for a free scalar field $f$ on
a fixed flat 2 dimensional spacetime $(M, \eta_{AB} )$ in terms of global inertial coordinates $X^A,\;A=0,1$ is 
\begin{equation}
S_0[f] = -\frac{1}{2} \int_M d^{2}X \eta^{AB}\partial_Af \partial_Bf ,
\label{s0}
\end{equation}
where  the Minkowski metric 
in inertial coordinates, $\eta_{AB}$,  is diagonal with entries $(-1,1)$. As is well known, solutions to the equations 
ensuing from the action (\ref{s0}) take the form 
$f= f^+(X+T) + f^-(T-X)$ where $f^{\pm}$ are arbitrary  functions of their `light cone' arguments. Due to this
functional dependence $f^+$ describes left moving modes and $f^-$, right moving modes on the flat spacetime.

If, in the action (\ref{s0}),  we use coordinates
$x^{\alpha}\;, \alpha= 0,1$ (so that $X^A$ are `parameterized' by $x^{\alpha}$,  $X^A= X^A(x^{\alpha})$), we have
\begin{equation}
S_0[f] = -\frac{1}{2} \int_M d^{2}x \sqrt{\eta} \eta^{\alpha \beta}\partial_{\alpha}f \partial_{\beta}f ,
\label{s0x}
\end{equation}
where $\eta_{\alpha \beta}= \eta_{AB} \partial_{\alpha}X^A\partial_{\beta}X^B$ and $\eta$ denotes the 
determinant of $\eta_{\alpha \beta}$. The action for PFT is obtained 
by considering the right hand side of (\ref{s0x}) as a functional, not only of $f$, but also of 
$X^A(x)$ i.e. $X^A(x)$ are considered as  2 new scalar fields to be varied in the action so that $\eta_{\alpha \beta}$
is considered to be a functional of $X^A(x)$). Thus 
\begin{equation}
S_{PFT} [f, X^A]= -\frac{1}{2} \int_M d^{2}x \sqrt{\eta(X)} \eta^{\alpha \beta}(X)\partial_{\alpha}f \partial_{\beta}f .
\label{spft}
\end{equation}
Note that $S_{PFT}$ is a diffeomorphism invariant functional of the scalar fields $f (x), X^A (x)$.
Variation of $f$ yields the equation of motion $\partial_{\alpha}(\sqrt{\eta}\eta^{\alpha \beta}\partial_{\beta}f)= 0$,
which is just the flat spacetime equation $\eta^{AB}\partial_A\partial_B f=0$ written in the coordinates $x^{\alpha}$.
On varying $X^A$, one obtains equations which are satisfied if  $\eta^{AB}\partial_A\partial_B f=0$.
This implies that $X^A(x)$ are   undetermined functions (subject to the condition that the 
determinant of $\partial_{\alpha}X^A$
is non- vanishing). This 2 functions- worth of gauge is a reflection of the 2 dimensional diffeomorphism invariance
of $S_{PFT}$. Clearly the dynamical content of $S_{PFT}$ is the same as that of $S_0$; it is only that 
the diffeomorphism invariance of 
$S_{PFT}$ naturally allows a description of the standard free field dynamics dictated by $S_0$ on {\em arbitrary}
foliations of the fixed flat spacetime.

The spacetime is assumed to be of topology $\Sigma \times R$.
A 1+1 Hamiltonian decomposition yields a phase space coordinatized by the canonically conjugate pairs
$(T(x),\Pi_T(x)),(X(x),\Pi_X(x) ), (f(x),\pi_f(x))$ where $x$ coordinatizes the 1 dimensional 
$t=$ constant
Cauchy slice $\Sigma$.
For each value of $x$, the functions $(T(x), X(x))$ locate a point in 
flat spacetime by virtue of their interpretation as Minkowskian coordinates so that as $x$ varies, $(T(x), X(x))$ 
describe an {\em embedding} of the Cauchy slice coordinatized by $x$ into the flat spacetime. As a result we 
refer to the sector of phase space
coordinatized by  $(T(x),\Pi_T(x)),(X(x),\Pi_X(x) )$ as the embedding sector.
A canonical transformation can be made into `left and right moving' embedding variables
$(X^+, \Pi_+), (X^, \Pi_-)$ with $X^{\pm}= T\pm X$ and $\Pi_{\pm}$ their conjugate momenta.
It is also useful to transform to the variables $Y^{\pm}= \pi_f + f^{\prime}$ in the matter sector. 
It can be checked that the `+' and `-' variables Poisson commute with each other.

The dynamics is generated by a pair of constraints 
\begin{equation}\label{eq:2}
{{H_{\pm}}}(x)\ =\ [\ \Pi_{\pm}(x)X^{\pm'}(x)\ \pm\
\frac{1}{4} (Y^{\pm}(x))^2     \ ].
\end{equation}
These constraints are of density weight two. In 1 spatial dimension their transformation properties under
coordinate transformations are identical to those of  
spatial covector fields. Integrating them against multipliers $N_{\pm}$, which can therefore be thought of as 
spatial vector fields, one finds that the integrated `+' (respectively `-')  constraint generates 
spatial diffeomorphisms on the 
`+' (respectively `-') variables while keeping the `-' (respectively `+')  variables untouched. Thus, PFT dynamics can be
thought of as the action of two {\em independent} spatial diffeomorphisms $\Phi^+, \Phi^-$ on the 
`+' and `-' sectors of the phase space.

Instead of the $H_{\pm}$ constraints we may consider the constraints 
\be
C_{diff} = H_+ +H_- 
\label{cdiff}
\ee
and 
\be
C_{ham}= H_+ - H_- .
\label{cham}
\ee
It can be checked that the diffeomorphism constraint $C_{diff}$ 
generates spatial diffeomorphisms on the Cauchy data $(X^-,\Pi_-, X^+,\Pi_+, Y^+,Y^-)$
whereas the (density weight two) Hamiltonian constraint $C_{ham}$ 
generates evolution of this data along the unit timelike normal $n^{\alpha}$ 
to the slice \cite{karel1} (recall that
the phase space data $(X^+,X^-)$ define an embedded Cauchy slice in Minkowski spacetime;  $n^{\alpha}$ 
is the  unit timelike normal to this slice with respect to the flat spacetime metric).

In terms of the finite transformations $\Phi_+, \Phi_-$, it follows from (\ref{cdiff}), (\ref{cham}) that 
finite spatial diffeomorphisms generated by $C_{diff}$  
correspond to the choice $\Phi_+ = \Phi_-$ whereas finite transformations generated
by the Hamiltonian constraint $C_{ham}$ correspond to the choice $\Phi_+ = (\Phi_-)^{-1}$.

The relation between evolution on phase space data generated by the constraints and free field evolution of $f$
on flat spacetime can be seen as follows \cite{karel1}.
The constraints ($C_{diff}, C_{ham}$ or, equivalently, $H_+,H_-$) generate transformations
of the phase space data $(X^{\pm}(x), Y^{\pm}(x))$ to new data $(X_{new}^{\pm}(x), Y_{new}^{\pm}(x))$.  
The phase space data $(X^+(x),X^-(x))$ define an embedded slice in flat spacetime.
Initial data on this slice for evolution  via the scalar wave equation can be given in terms of 
left and right moving values of the scalar field on the slice (or, ignoring issues of zero modes, the values of 
their derivatives). As discussed in \cite{karel1} the relationship between the phase space data $Y^{\pm}$ and these
derivatives is given by 
$\frac{\partial f}{\partial X^{\pm}}|_{X^{\pm}= X^{\pm}(x)}= \frac{Y^{\pm}(x)}{X^{\pm \prime}(x)}$. 
The transformed embedding data $X^{\pm}_{new}$ then 
define  an `evolved' slice in flat 
spacetime  with  matter phase space data 
${X_{new}^{\pm \prime}(x)}\frac{\partial f}{\partial X^{\pm}}|_{X^{\pm}= X_{new}^{\pm}(x)}= {Y_{new}^{\pm}(x)}$
where $f$ is the restriction to the new slice of the solution  to the wave equation with initial data
$\frac{\partial f}{\partial X^{\pm}}$ on the old slice.

As in the previous works \cite{polypft1,polypft2,polypft3} we shall restrict attention to a flat spacetime $(M,\eta )$ with 
cylindrical topology $S^1 \times R$. We denote the length of the $T=0$ circle in the flat spacetime $(M,\eta )$ by $L$. 
The cylindrical topology of $M$ implies that any 
Cauchy slice $\Sigma$ is  circular.
Certain subtelities related to  the use of a single  angular inertial coordinate $X$ on the flat spacetime $M$
as well as a single angular coordinate $x$  
on the Cauchy  slice $\Sigma$ arise but these subtelities and their ramifications constitute   technical 
details which
may be ignored in as much as the  key arguements 
in this work are concerned. 
The interested reader may consult \cite{karel1} and section 
IIC of \cite{polypft1} for an account of these subtelities in classical theory.

\subsection{Quantum Theory\label{sqt}}

We shall concentrate mainly on the embedding sector in this brief review. 
For further details regarding the matter sector please see \cite{polypft2,polypft3}.
We shall mention, but not explain in any detail, the  subtelities in the quantum theory concerning the circular spatial
topology because such details will only serve to distract from the main arguments in subsequent sections. 
The interested reader may consult \cite{polypft2} for such details.

\subsubsection{Kinematics\label{sqtk}}

The embedding sector Hilbert space is a tensor product of  `+' and `-' sectors. 
On the `$+$' sector the operator 
correspondents of functions on the `-' sector of phase space act trivially and vice versa.

The `+' embedding sector 
is spanned by an orthonormal  basis of  charge network states each of which is 
denoted by $|\gamma_{+}, \vec{ k^{+}} \ket$ where $\gamma_{+}$ is a graph i.e.
a set of edges which cover the circle with  each edge $e_{}$ labelled by a `charge' $k^{+}_{e}$, 
the collection of  such charges for all the edges in the graph being denoted by $\vec{k^{+}}$.
The $+$ embedding sector Hilbert space 
provides a representation of the Poisson bracket algebra between the classical `holonomy functions'
$\exp i(   \sum_{e} ( k_{e}^{+}\int_{e_{}}\Pi_+))$, and the embedding coordinate $X^{+}(x)$. 
In this representation the embedding momenta are polymerized so that  the holonomy functions 
are well defined operators but the embedding momenta themselves are not. The embedding coordinate
operator ${\hat X}^+(x)$ is well defined and the charge net states are eigen states of this operator.
In particular the action of the embedding coordinate operators ${\hat X}^{+}(x)$ on a charge network state
$|\gamma_{+}, \vec{k^{+}}\ket$
when the argument $x$ lies in the interior of an edge $e$ of $\gamma_+$ is:
\be
{\hat X}^{+}(x) |\gamma , \vec{k^{+}}\ket = \hbar k_e^{+} |\gamma , \vec{k^{+}}\ket 
\label{xplushat}
\ee
Identical results hold for $+\rightarrow -$. 
The  tensor product states $|\gamma_{+}, \vec{k^{+}}\ket\otimes |\gamma_{-}, \vec{k^{-}}\ket$ form a basis of the 
embedding sector Hilbert space and are referred to as embedding charge network states.
By going to a graph finer than $\gamma_+, \gamma_-$, each such state can be equally well denoted by 
$|\gamma, \vec{k^+}, \vec{k^-} \ket$ where each edge $e$  of the graph $\gamma$ is labelled by 
a {\em pair} of charges $(k^+_e, k^-_e)$ and the collection of these charges is denoted by $(\vec{k^+}, \vec{k^-})$.
Such a state is an eigen ket of both the ${\hat X}^+$ and ${\hat X}^-$ operators.
Similar to (\ref{xplushat}) the action of ${\hat X}^{\pm}(x)$ on the 
charge net $|\gamma, \vec{k^+}, \vec{k^-} \ket$ when $x$ is in the interior of  an edge $e$ of $\gamma$ is:
\be
{\hat X}^{\pm}(x) |\gamma , \vec{k^{+}},  \vec{k^{-}}\ket = \hbar k_e^{\pm} |\gamma , \vec{k^{+}},  \vec{k^{-}}\ket 
\label{xspect}
\ee
The charges are chosen to be integer valued multiples of a dimensionful parameter $\frac{a}{\hbar}$ 
so that 
\be
\hbar k^{\pm}_e \in {\bf Z}a
\label{xspect1}
\ee
 where $a$ is a Barbero- Immirzi like parameter. In the context of the circular spatial topology
relevant to this work, we restrict attenion, as in \cite{polypft2}, to a value of $a$ such that 
\be
\frac{L}{a} = N,  
\label{defn}
\ee
$N$ being a positive integer.


The matter sector Hilbert space is also a tensor product of  `+' and `-' sectors. 
On the `+' sector the field $Y^{+}$ is polymerised and on the `-' sector the  $Y^-$ field is polymerised.
Thus taken together, neither $Y^+$ nor $Y^-$ (and hence neither $\pi_f$ nor $f$) exist as well defined operators.
The `+' sector provides 
a representation of matter holonomy functions
$\exp i(\sum_e l_e^+\int_eY_+ )$ on a basis of `+' matter charge nets, each such charge net denoted by 
$|\gamma_+, \vec{l^+}\ket$ in obvious notation. The matter charges ${\vec l^+}$ on each such charge net are real
and are subject to the following  restriction \cite{polypft2}: for every pair of edges
edges $e,e' \in \gamma_+$, the matter charge difference $l^+_e - l^+_{e'}$ is an integer valued multiple of a dimensionful
parameter ${\bf \epsilon}$.
\footnote{We do not impose the  vanishing zero constraint \cite{polypft1,polypft2}. We remark further on the
zero mode issue in section \ref{sdiscuss}.}
The `-' sector structure is identical. The parameter 
${\bf \epsilon}$ along with the parameter `$a$' for the 
embedding sector constitute `Barbero-Immirzi' parameters and label inequivalent representations.
The  tensor product states $|\gamma_{+}, \vec{l^{+}}\ket\otimes  |\gamma_{-}, \vec{l^{-}}\ket$ form an orthonormal
 basis of the  matter Hilbert space and are referred to as matter charge net states.
By going to a fine enough graph any such state can be denoted, in notation similar to that for embedding states, 
as $|\gamma, \vec{l^+}, \vec{l^-} \ket$. 

The tensor product of the matter and embedding Hilbert spaces yields the kinematic Hilbert space ${\cal H}_{kin}$ for PFT.
This Hilbert space is spanned by charge net states each of which is a tensor product of a matter charge net 
and an embedding charge net. By going to a fine enough graph underlying the matter and embedding charge nets 
we may denote such a tensor product state by 
$|\gamma, \vec{k^+}, \vec{k^-}, \vec{l^+}, \vec{l^-} \ket$.

Since the `+' and `-' sectors are independent, we also have the tensor product  decomposition:
\be
|\gamma, \vec{k^+}, \vec{k^-}, \vec{l^+}, \vec{l^-} \ket =  |\gamma_+, \vec{k^+}, \vec{l^+}\ket \otimes
|\gamma_-, \vec{k^-}, \vec{l^-} \ket
\label{ptn}
\ee
  where 
$|\gamma_{\pm}, \vec{k^{\pm}}, \vec{l^{\pm}}\ket$ is itself a 
product of a `$\pm$' embedding charge network and a `$\pm$' matter charge  network.

\subsubsection{Gauge transformations generated by the constraints.\label{sg}}

Recall that the finite transformations generated by the  constraints $H_+, H_-$ correspond to a pair
of diffeomorphisms $\Phi_+, \Phi_-$. The quantum kinematics supports a unitary  representation of 
these diffeomorphisms by the unitary operators ${\hat U}_+(\Phi_+ ), {\hat U}_-(\Phi_- )$.
The operator ${\hat U}_+(\Phi_+ )$ acts on a `+' charge network state 
$|\gamma_+, \vec{k^+}, \vec{l^+}\ket$  by moving the graph and its colored edges by the diffeomorphism
$\Phi_+$ while acting as identity on `-' charge network states, and a similar action holds  for $+\rightarrow -$.
\footnote{While the matter charges are unaffected, each time an edge moves past the point $x=0\equiv x=2\pi$
 its embedding charge labels are augmented by factors of $L$ where $L$ is the length of the $T=$constant circles 
in flat spacetime. This is one of the subtelities arising from circular spatial topology.
\label{fn1}}
We denote this action by 
\be
{\hat U}_{\pm}(\Phi_{\pm})|\gamma_{\pm}, \vec{k^{\pm}}, \vec{l^{\pm}}\ket 
=: |\gamma_{\pm, \Phi_{\pm}}, \vec{k^{\pm}_{\Phi_{\pm}}}, \vec{l^{\pm}_{\Phi_{\pm}}}\ket
\label{uphipm}
\ee
The action of finite gauge transformations on a  charge net state 
$|\gamma, \vec{k^+}, \vec{k^-}, \vec{l^+}, \vec{l^-} \ket$ can then be deduced from equation (\ref{ptn}):
\be
{\hat U}_{+}(\Phi_{+}) {\hat U}_{-}(\Phi_{-})|\gamma, \vec{k^+}, \vec{k^-}, \vec{l^+}, \vec{l^-} \ket
= |\gamma_{+, \Phi_{+}}, \vec{k^{+}_{\Phi_{+}}}, \vec{ l^{+}_{ \Phi_{+} }}\ket \otimes
|\gamma_{-, \Phi_{-}}, \vec{k^{-}_{\Phi_{-}}}, \vec{l^{-}_{\Phi_{-}}}\ket .
\label{uphipphim}
\ee
By going to a graph finer than   $\gamma_{+, \Phi_{+}},  \gamma_{_, \Phi_{-}}$ the right hand side can 
again be written as a chargenet labelled by a single graph with each edge  labelled by a set of 4 charges
namely  the `+' and `-' embedding and matter charge labels.

\subsubsection{Group Averaging \label{sga}}

Physical states can be constructed by group averaging \cite{grpavge}  over the action of all $\Phi_+, \Phi_-$.
From \cite{polypft2} we have that the group average of  any charge net 
$|\gamma, \vec{k^+}, \vec{k^-}, \vec{l^+}, \vec{l^-} \ket :=|\gamma_+, \vec{k^+}, \vec{l^+}\ket \otimes
|\gamma_-, \vec{k^-}, \vec{l^-} \ket $ is the distribution $\Psi$ given by:
\ba
\Psi &:= &\sum_{(\Phi_+, \Phi_-)\in {\rm Orbit}} 
\bra\gamma, \vec{k^+}, \vec{k^-}, \vec{l^+}, \vec{l^-} |{\hat U}^{\dagger}_{+}(\Phi_{+}) {\hat U}^{\dagger}_{-}(\Phi_{-})
\nonumber
\\
& =& \sum_{(\Phi_+, \Phi_-)\in {\rm Orbit}}
\bra \gamma_{+, \Phi_{+}}, \vec{k^{+}_{\Phi_{+}}}, \vec{ l^{+}_{ \Phi_{+} }}| \otimes
\bra \gamma_{-, \Phi_{-}}, \vec{k^{-}_{\Phi_{-}}}, \vec{l^{-}_{\Phi_{-}}} | 
\label{grpavge}
\ea
where ${\rm Orbit}$ comprises of gauge transformations such that for each distinct chargenet which is gauge related
to $|\gamma, \vec{k^+}, \vec{k^-}, \vec{l^+}, \vec{l^-} \ket$ there is a unique element $(\Phi_+, \Phi_-)$
in ${\rm Orbit}$ which maps  $|\gamma, \vec{k^+}, \vec{k^-}, \vec{l^+}, \vec{l^-} \ket$  to this gauge related
image. In other words, the sum is over all distinct gauge related chargenets. We shall ignore
issues of ambiguities in the group averaging procedure (see for example \cite{alm2t,polypft1,polypft2})
 as this will not be important for the arguements in this  paper.

\section{The Finest Lattice Sector and its Pictorial Representation \label{sec3}}

Since the embedding charges are {\em eigen values} of the embedding coordinate operators, we can associate 
an embedding charge net with a discrete slice of Minkowski spacetime as follows.
For every edge $e$ the pair $\hbar k^+_e, \hbar k^-_e$ specifies  the point $X^+= \hbar k^+_e, X^-=\hbar k^-_e$
in flat spacetime.
The set of such points for all edgewise pairs of eigen values then defines a set of points in the 
flat spacetime which we may refer to as a discrete slice.
It turns out \cite{polypft2,polypft3} 
that there exists a superselected sector (with respect to  all gauge transformations together with 
a complete set of Dirac observables) of states with the following `finest lattice' property. 

Consider a lightcone lattice of spacing $a$ in the flat spacetime so that $X^+,X^- \in {\bf Z}a$ on this lattice.
We shall say that a pair of points are nearest neighbours if they are either light like seperated and one lattice
spacing away from each other 
or if they are spacelike seperated and located at a spatial distance $a$ (as measured by the flat spacetime
metric) from each other. Thus each point on the lattice has 6 nearest neighbours, 4 lightlike and 2 spacelike.
\footnote{There are also 2 `timelike' neighbours. It turns out that these are not relevant to our discussion. Hence
we exclude them from our definition of nearest neighbours.
}
Next, consider the set of edgewise pairs of embedding charges for any charge net in the superselected sector
under discussion. Plot these  as a set of points in flat spacetime in the manner described above.
Then it turns out that these points fall on the spacetime lattice  $X^+,X^- \in {\bf Z}a$ in such a way that 
points obtained from adjacent edges are nearest neighbours. 
Further, by virtue of the minimal spacing of the 
eigen values of the $X^{\pm}$ operators (\ref{xspect})- (\ref{xspect1}), 
flat spacetime point sets defined by {\em any} chargenet in the 
kinematic Hilbert space  cannot fall on any finer lattice.
Finally, for any charge net in this sector (see (i) below), the matter charges are distributed on the underlying 
graph in a coarser manner than the embedding charges. Hence with each pair of embedding charges which specify 
a point in flat spacetime, 
we can {\em uniquely} associate a pair of matter charges. This means that we
can label each point on the discrete slice in flat spacetime 
 by a pair of matter charges so that the chargenet can be interpreted
as a specification of quantum matter on a discrete `Cauchy' slice, this slice satisfying the `nearest neighbour' property
on the finest lattice allowed by the spectrum (\ref{xspect}),(\ref{xspect1}). 
No other chargenet outside this superselection sector
admits this interpretation. Hence this sector is called the `finest lattice' sector.

Technically, chargenets in this sector are specified as follows \cite{polypft3}. A chargenet 
$|\gamma, \vec{k^+}, \vec{k^-}, \vec{l^+}, \vec{l^-} \ket= |\gamma_+, \vec{k^+}, \vec{l^+}\ket \otimes
|\gamma_-, \vec{k^-}, \vec{l^-} \ket$ belongs to the finest lattice sector iff (i)- (iii) below hold:
\\

(i) its  matter chargenet labels are `coarser' than the embedding ones so that each pair of successive edges
 of the coarsest graph
$\gamma^{coarse}_{\pm}$ underlying $|\gamma_{\pm}, \vec{k^{\pm}}, \vec{l^{\pm}} \ket$ is necessarily labelled
by distinct pairs of $\pm$ embedding charges but not necessarily distinct pairs of $\pm$ matter  charges.\\

(ii) its  embedding charges on the coarsest graph $\gamma^{coarse}_{\pm}$ underlying 
$|\gamma_{\pm}, \vec{k^{\pm}}, \vec{l^{\pm}} \ket$
 satisfy $\hbar k^{\pm}_{e^{\prime}_{\pm}}-  \hbar k^{\pm}_{e_{\pm}} = a$
where  $e^{\prime}_{\pm},e_{\pm}$ are adjacent edges in $\gamma^{coarse}_{\pm}$ such that 
$e^{\prime}_{\pm}$
lies to the right of $e_{\pm}$  in the coordinatization  $x$ 
(i.e.the edges are located such that given any point $p_{\pm}$ in the interior of  $e_{\pm}$ 
with coordinate $x=x_{\pm}$ and any point 
$p_{\pm}^{\prime}$ in the interior of  $e_{\pm}^{\prime} $ with coordinate 
$x=x_{\pm}^{\prime}$, we have that  $x_{\pm}^{\prime} > x_{\pm}$).

Further,
 if the difference in the embedding charge value on  the last edge and  the first edge of 
$\gamma^{coarse}_{\pm}$ is $\pm\frac{L}{\hbar}$, the matter charge values on these  edges are 
identical.\footnote{This additional restriction on the matter charges is due to subtelities connected with circular
spatial topology (see Footnote \ref{fn1}).}
\\

(iii) there are $N_+$ `+' and $N_-$ `-' distinct embedding charges with $N_+$ such that either $N_+=N+1$
or $N_+=N$  
and $N_-$ such that either  $N_-= N+1$
or $N_-= N$ where $N$ is defined in equation (\ref{defn}).
\footnote{\label{fnN}We assume that $N>>4$ for certain technical reasons (related to the 
circular spatial topology) in connection with proofs in the following sections and in the Appendix.}
\\

Property (i) ensures that we may think of the matter charges $l^+_e,l^-_e$ as sitting
on the lattice point $\hbar k^+_e, \hbar k^-_e$. Property (ii) is 
 the technical formulation of  the `nearest neighbour'
condition. Property (iii) turns out to be necessary for the consistency of the `discrete Cauchy slice' interpretation 
in the context of circular spatial topology.
To summarise: The kinematic charge nets of polymer PFT in this sector can be interpreted as
describing quantum matter degrees of freedom on discrete Cauchy slices which fall on a (light cone) lattice
in Minkowski spacetime.  
For the remainder of this work we shall focus, for concreteness,
exclusively on states in this finest lattice sector.

Let us consider the action of a gauge transformation $\Phi_+$ on a finest lattice state
$|\gamma, \vec{k^+}, \vec{k^-}, \vec{l^+}, \vec{l^-} \ket$. It moves the `+' charged edges along the circle
relative to the fixed\\ `-' charged edges. Consider a fine enough graph which underlies 
the new set of `+' edges together with the old `-' edges. On this graph the set of {\em edgewise pairs} of charges
is different from the set corresponding to the original charge net. This new set defines a {\em new} discrete
Cauchy slice and matter data on this slice. Thus, the matter data {\em propagate} from one discrete Cauchy slice to 
another.  A similar picture ensues for the action of a gauge transformation
$\Phi_-$ and one explicitly sees how, just as in classical theory, the finite gauge transformations generated
by the constraints propagate matter from one slice to another. It turns out 
(see \cite{polypft2} for details) that 
the considerations of section \ref{sg} in conjunction with Footnote \ref{fn1} ensure that,  
for appropriate choices of gauge 
transformations, the initial and final discrete slices can be `macroscopically' seperated 
(i.e. by arbitarily many
lattice points) in time so that such gauge transformations
implement  {\em long range propagation}.

From the discussion in section \ref{sga} a physical state obtained by group averaging over a charge net 
$|\gamma, \vec{k^+}, \vec{k^-}, \vec{l^+}, \vec{l^-} \ket$ can be written as a sum over all distinct 
charge net states obtained from this one by action of all possible finite gauge transformations. As asserted 
in \cite{polypft2}, if we plot the lattice points in flat spacetime associated with each of these states together
with their matter charge labelling, one obtains a single discrete spacetime lattice with uniquely specified
matter charges at each lattice point. This specification is {\em consistent} in the sense that if a single
lattice point derives from  the same  pair of embedding charges arising from different states in the sum,
the matter charge labels for this point provided by these different states are the same. In other words the 
discrete Cauchy slices with matter data which occur as summands in (\ref{grpavge}) stack up consistently to give
a single discrete spacetime with quantum matter at each spacetime point.
Putting this picture together
with that of the action of finite gauge transformations discussed above it follows that 
any physical state  corresponding to a group averaged finest lattice charge net 
encodes quantum matter propagation on a discrete spacetime.

Before we conclude this section, we define some useful terminology. First note that {\em edges} of a chargenet
define {\em points} in flat spacetime.  If two successive edges define light like related points we refer to the 
vertex of the chargenet at which these edges intersect as a {\em null} vertex. If the points are spacelike related
we refer to the corresponding vertex as a {\em spacelike} vertex. Note that from (ii) above successive edges 
can never define flat spacetime points which are timelike seperated.

\section{Ultralocality of the Hamiltonian constraint and Smolin's criticism \label{sec4}}

In the section \ref{sga} we showed that physical states could be constructed by group averaging over the 
finite transformations generated by the  `light cone' constraints $H_+, H_-$. Since Smolin's criticisms
apply to the LQG formalism wherein finite spatial diffeomorphisms are averaged over and the Hamiltonian constraint
is constructed via Thiemann's procedure, we now turn our attention to a similar treatment for PFT in terms
of its diffeomorphism and Hamiltonian constraints, $C_{diff}$(\ref{cdiff})   and $C_{ham}$ (\ref{cham})

First consider, as in LQG, the finite transformations generated by $C_{diff}$.
Recall from the discussion after (\ref{cdiff}), (\ref{cham}) that these transformations correspond to the case $\Phi_+=\Phi_-$.
Indeed, setting $\Phi_+=\Phi_-= \Phi$ in equation (\ref{uphipphim})we see that the associated unitary transformations 
${\hat U}_+(\Phi) {\hat U}_-(\Phi) $ drag {\em both} the `+' and `-' chargenets by the same diffeomorphism
$\Phi$, which of course, just corresponds to the action of a spatial diffeomorphism
\footnote{This turns out to be true despite the added subtelities alluded to in Footnote \ref{fn1} 
\cite{polypft1,polypft2}. 
}:
\ba
{\hat U}_{diff}(\Phi )|\gamma, \vec{k^+}, \vec{k^-}, \vec{l^+}, \vec{l^-} \ket &=&
{\hat U}_{+}(\Phi) {\hat U}_{-}(\Phi)|\gamma, \vec{k^+}, \vec{k^-}, \vec{l^+}, \vec{l^-} \ket \nonumber\\
&=& |\gamma_{+, \Phi}, \vec{k^{+}_{\Phi_{}}}, \vec{ l^{+}_{ \Phi_{} }}\ket \otimes
|\gamma_{-, \Phi_{}}, \vec{k^{-}_{\Phi_{}}}, \vec{l^{-}_{\Phi_{}}}\ket .
\label{udiff}
\ea
By going to  a finer graph that $\gamma_{+, \Phi},\gamma_{-, \Phi_{}}$, one can denote the tensor product 
chargenet  in the last line of the above equation by a single chargenet with edges labelled by 
a quadruple of (2 embedding and 2 matter) charges.
Since both the `+' and `-' labels are dragged around by the same diffeomorphism, it is straightforward to check
that ${\hat U}_{diff}(\Phi )|\gamma, \vec{k^+}, \vec{k^-}, \vec{l^+}, \vec{l^-} \ket$ and 
$|\gamma, \vec{k^+}, \vec{k^-}, \vec{l^+}, \vec{l^-} \ket$ define the same discrete Cauchy slice in flat spacetime with
the same matter data.  This is as it should be because the flat spacetime picture  encodes the {\em relation}
between the embedding and matter excitations and this relation is  diffeomorphism invariant.
Averaging over diffeomorphisms can then be done \cite{polypft3}. Specialising (\ref{grpavge}) to
the case $\Phi_+= \Phi_-= \Phi$, and ignoring group averaging ambiguities, we have that 
\ba
\Psi_{diff} &:= &\sum_{\Phi \in {\rm Orbit_{diff}}} 
\bra\gamma, \vec{k^+}, \vec{k^-}, \vec{l^+}, \vec{l^-} |{\hat U}^{\dagger}_{+}(\Phi_{}) {\hat U}^{\dagger}_{-}(\Phi_{})
\nonumber
\\
& =& \sum_{\Phi \in {\rm Orbit_{diff}}}\bra \gamma_{+, \Phi_{}}, \vec{k^{+}_{\Phi_{}}}, \vec{ l^{+}_{ \Phi_{} }}| \otimes
\bra \gamma_{-, \Phi_{}}, \vec{k^{-}_{\Phi_{}}}, \vec{l^{-}_{\Phi_{}}} | 
\label{diffavge}
\ea
where ${\rm Orbit_{diff}}$ consists of elements which uniquely take the charge net being averaged over to 
its distinct diffeomorphic images.
Thus the  
distribution $\Psi_{diff}$ is a sum over distinct diffeomorphic images of the chargenet being averaged so that 
if a certain bra is in the sum, then all its diffeomorphic images are also in the sum.

Next consider the Hamiltonian constraint  $C_{ham}$ (\ref{cham}). 
Recall from 
discussion after (\ref{cdiff})-(\ref{cham}) that the finite 
transformations generated by $C_{ham}$ correspond to the case $\Phi_+=(\Phi_- )^{-1}$.
From Reference \cite{polypft3}, it follows that with particular quantization choices 
\footnote{We shall comment more on these choices in Section \ref{sdiscuss}. \label{fnhamchoice}}
at finite triangulation,
the finite triangulation approximant to the Hamiltonian constraint $C_{ham}(N)$ smeared with lapse $N$ can be written as 
\ba
{\hat C}_{ham,\delta}(N)|\gamma, \vec{k^+}, \vec{k^-}, \vec{l^+}, \vec{l^-} \ket
&=&-i\hbar \sum_{v} N(v) \frac{{\hat U}_+(\Phi_{\delta, v} ){\hat U}_-(\Phi^{-1}_{\delta, v} )- {\bf 1}}{\delta}
|\gamma, \vec{k^+}, \vec{k^-}, \vec{l^+}, \vec{l^-} \ket \\
&=& -i\hbar \sum_{v} N(v)
|\gamma_{+, \Phi_{\delta}}, \vec{k^{+}_{\Phi_{\delta}}}, \vec{ l^{+}_{ \Phi_{\delta} }}\ket \otimes
|\gamma_{-, \Phi^{-1}_{\delta}}, \vec{k^{-}_{\Phi^{-1}_{\delta}}}, \vec{l^{-}_{\Phi^{-1}_{\delta}}}\ket
\label{chat0}
\ea
The sum  in (\ref{chat0}) is over  `non-trivial' vertices of the (finest lattice) charge net 
$|\gamma, \vec{k^+}, \vec{k^-}, \vec{l^+}, \vec{l^-} \ket$. By a `non-trivial' 
vertex we mean
a point on $\gamma$  at which the ingoing and outgoing edges carry non-identical charges. Since the charge net is 
a finest lattice charge net, this implies that at least one of the incoming embedding charges differs from
its outgoing counterpart. $\Phi_{\delta, v}$ is a
small diffeomorphism around the non-trivial vertex $v$ which 
moves  $v$ to its right by an amount $\delta$ as measured by the
coordinate $x$, where `right' means direction of increasing $x$. In addition $\Phi_{\delta , v}$ is identity
outside a region of size of order $\delta$ around $v$ and $\Phi_{\delta , v}$ is such that its inverse
$\Phi^{-1}_{\delta , v}$ moves $v$ to its left by an amount $\delta$.
Thus given a chargenet, the Hamiltonian 
constraint acts only at its non-trivial vertices and moves the `+' part of the charge net in the vicinity of 
the vertex $v$
to the right and the `-' part of the chargenet in the vicinity of the vertex $v$ to the left. 
For small enough $\delta$ it is easy to see that the diffeomorphism class of the chargenet deformed 
in this manner remains the same as $\delta\rightarrow 0$. Finally note  that we may extend the sum 
in (\ref{chat0}) to include any number of  `trivial' vertices of $\gamma$
for which the incoming and outgoing charges are identical because, as is straightforward to verify, on any such 
vertex $v$, the operator ${\hat U}_+(\Phi_{\delta, v} ){\hat U}_-(\Phi^{-1}_{\delta, v})$ acts as the identity
on the chargenet so that these vertices do not yield non-zero contributions.
It is convenient to define:
\be
{\hat U}_{ham, \delta, v}:= {\hat U}_+(\Phi_{\delta, v} ){\hat U}_-(\Phi^{-1}_{\delta, v}).
\label{defuham}
\ee
so that 
\be 
{\hat C}_{ham,\delta}(N)|\gamma, \vec{k^+}, \vec{k^-}, \vec{l^+}, \vec{l^-} \ket
= -i\hbar \sum_v N(v)\frac{ {\hat U}_{ham, \delta, v} -{\bf 1} }{\delta }
|\gamma, \vec{k^+}, \vec{k^-}, \vec{l^+}, \vec{l^-} \ket
\label{chat}
\ee
where it is understood that the sum ranges over all non-trivial vertices but can also include 
any number of trivial vertices as well.

Next, we note that choices could equally well be made in the Thiemann procedure 
\footnote{We discuss this further in Section \ref{sdiscuss}.}
 such that we obtain
the action of the finite triangulation constraint to be 
\be 
{\hat C}_{ham,\delta}(N)|\gamma, \vec{k^+}, \vec{k^-}, \vec{l^+}, \vec{l^-} \ket
= i\hbar \sum_v N(v)\frac{ {\hat U}_{ham, \delta , v}^{\dagger} -{\bf 1} }{\delta }
|\gamma, \vec{k^+}, \vec{k^-}, \vec{l^+}, \vec{l^-} \ket
\label{chata}
\ee
where, from equation (\ref{defuham}), we have that
\be
{\hat U}^{\dagger}_{ham, \delta, v}= {\hat U}^{\dagger}_-(\Phi^{-1}_{\delta, v}){\hat U}^{\dagger}_+(\Phi_{\delta, v} )
= {\hat U}_-(\Phi_{\delta, v}){\hat U}_+(\Phi^{-1}_{\delta, v} )
={\hat U}_+(\Phi^{-1}_{\delta, v} ){\hat U}_-(\Phi_{\delta, v})
\label{defuhama}
\ee
The action of the finite triangulation Hamiltonian constraint (\ref{chata}) would then be to move 
the `+' part of the charge net in the vicinity of the vertex $v$
to the {\em left}  and the `-' part of the chargenet in the vicinity of the vertex to the {\em right}. 
Clearly the  actions (\ref{chat}), (\ref{chata}) are ultralocal in the sense of Smolin. More in detail, 
these actions are  only in the vicinity of  vertices of the chargenet being acted upon and
the action at one vertex is completely independent of the other. 
We now analyse repeated actions of the Hamiltonian constraint and show that they do not lead to propagation.

We shall focus first on the action of ${\hat U}_{ham, \delta, v}$ at a  
vertex $v$ with  incoming embedding charges $(k_1^+,k_1^-)$ and outgoing charges $(k_2^+, k_2^-)$.
The action bifurcates the vertex, giving rise to two new vertices $v_1, v_2$ with $v_1$ to the left of $v_2$ 
and one new edge connecting these vertices
 so that the sequence of charges from left to right  is now $(k_1^+,k_1^-),(k_1^+,k_2^-),(k_2^+,k_2^-)$. Thus
the sequence of charges in the vicinity of $v$  changes as:
\be
(k_1^+,k_1^-), (k_2^+,k_2^-)\rightarrow (k_1^+,k_1^-),(k_1^+,k_2^-),(k_2^+,k_2^-)
\label{seq12}
\ee
The following three cases  are of interest 
(see the end of section \ref{sec3} for the terminology used in (ii),(iii) below):\\
(i) The original vertex $v$ is trivial: In this case the charge net is unchanged.\\
(ii)The vertex $v$ is {\em null} so that $k_1^+= k_2^+$ or $k_1^-=k_2^-$: In this case the 
new charge sequence is equivalent to the one on  the original charge net so that the new chargenet
is diffeomorphic to the old one. It follows that  the new charge net defines exactly the same points in 
flat spacetime as the old charge net. \\
(iii)The vertex $v$ is {\em spacelike} so that $k_1^+\neq k_2^+$ and  $k_1^-\neq k_2^-$. In this case
we see that the new vertices $v_1$ and $v_2$ are null. Further from (ii), section \ref{sec3} and (\ref{seq12}) it follows
that 
the new chargenet represents `one lattice step' of 
evolution with respect to the original chargenet.\\

Exactly the same conclusions ensue for the action (\ref{chata}). Note that the matter charges are just dragged
along together with the embedding charges by the actions (\ref{chat}),(\ref{chata}).
From the above discussion we see that if $v$ is trivial or null the Hamiltonian constraint actions
(\ref{chat}),(\ref{chata}) do not change the flat spacetime points (and the matter data thereon) 
associated with the chargenet so that there is no evolution in the flat spacetime.
If $v$ is spacelike, it follows from (iii) above that  the action of the Hamiltonian constraint (\ref{chat}),(\ref{chata})
evolves matter data by one lattice step. Note however that this action replaces the spacelike vertex $v$ of the charge 
net by a pair of null vertices  $v_1$ and $v_2$. From (ii) above it follows that {\em further} actions of 
the Hamiltonian constraint (whether (\ref{chat}) or (\ref{chata})) at these new vertices of the chargenet
 do not evolve  the discrete Cauchy slice  any further.

Applying these results to all the vertices of the chargenet we conclude that 
repeated  actions of the Hamiltonian constraint  ( say $n_1$ actions (\ref{chat}) followed by $m_1$ actions 
(\ref{chata}) followed by $n_2$ of (\ref{chat}) and so on)
cannot evolve the discrete Cauchy slice with quantum matter
any further than one lattice step away in flat spacetime. It is in this precise sense that the Hamiltonian 
constraint does not generate long range propagation and it is in this precise sense 
that  Smolin's criticism  is  formulated in the case of  PFT.

In order to understand exactly why repeated actions of the Hamiltonian constraint fail to generate long range 
propagation, it is appropriate to compare the actions (\ref{chat}), (\ref{chata}) with that of  {\em finite}
gauge transformations which, as explained in section \ref{sec3} and \cite{polypft2}, {\em do} describe
long range propagation. Let us consider 3 successive edges  $e_3,e_1,e_2$ of the charge net so that the embedding charge 
sequence is now $(k_3^+,k_3^-),(k_1^+,k_1^-),(k_2^+,k_2^-)$. Let us call the succesive vertices $u$ and $v$ so that 
$u$ is the `3-1' vertex and  $v$, as before, is the `1-2' vertex and let us assume that these vertices are spacelike. 
Now let us consider a sequence of actions of `-' diffeomorphisms each of 
which stretches the `-' labels to the left and each of which is identity to the left of $e_3$ and to the right of $e_2$. 
In visualising this action recall that the charge network has  a $`+ \otimes -$ product structure (\ref{ptn}).
With an appropriate choice of sequence of such `-' diffeomorphisms, we obtain the 
following sequence of embedding charges:
\ba
(k_3^+,k_3^-),(k_1^+,k_1^-),(k_2^+,k_2^-)
\rightarrow (k_3^+,k_3^-),(k_1^+,k_1^-),(k_1^+,k_2^-), (k_2^+,k_2^-) \rightarrow\;\;\;\; \label{4.10}\\
(k_3^+,k_3^-), (k_3^+,k_1^-),(k_1^+,k_1^-),(k_1^+,k_2^-), (k_2^+,k_2^-)
\rightarrow (k_3^+,k_3^-), (k_3^+,k_1^-),(k_1^+,k_2^-), (k_2^+,k_2^-) \;\;\;\;\label{no1}\\
\rightarrow (k_3^+,k_3^-), (k_3^+,k_1^-), (k_3^+,k_2^-),  (k_1^+,k_2^-), (k_2^+,k_2^-)
\;\;\;\;\;\;\;\;\;\;\;\;\;\;\; & \label{2step}
\label{seqphi}
\ea
Thus, at the end of this sequence of  actions, 
the  `-' edge with charge `$k_2^-$'  has moved leftward so as to overlap the
$k_3^+$ edge yielding the point $(\hbar k^+_3, \hbar k_2^-)$ which  is {\em 2 lattice spacings}  away
from points on the original slice. In order for this to happen in a continuous manner as depicted in 
(\ref{4.10})- (\ref{2step}), it is {\em necessary} to have the intermediate step
(\ref{no1}) where the $k_2^-$ charge completely displaces the $k_1^-$ charge from the `+' edge with label $k_1^+$.

In contrast, since the action of the Hamiltonian constraint
(\ref{chat}),(\ref{chata}) is for sufficiently small $\delta$, this action even if repeated, can never completely
erase any of the original pairs of edge labels. In particular even after repeated actions of the Hamiltonian 
constraint, there is always an edge with the label $(k_1^+,k_1^-)$. This survival of the original edge labels
is directly tied to the ultralocality of the action of the Hamiltonian constraint: since its action 
only modifies the structure in a small enough
\footnote{The precise notion of `small enough' is defined in {\bf L1}, section \ref{sec5}}  
 neighbourhood of a vertex, each one of the  original edges always has
a part which is not affected by this action.
Viewed in this manner, it is the {\em inability} of repeated actions of the Hamiltonian constraint to erase
such original pairs of labels which obstructs long range propagation.

As mentioned in  section \ref{sec1}, in the next section we offer a new perspective on propagation and show
how this obstruction is sidestepped. Once this obstruction is sidestepped it is not difficult to prove
that long range propagation ensues. Since we believe that this proof is just an added layer of polymer PFT 
technicalities, we relegate this proof to the appendix and concentrate in the main body of the paper
on the key lesson of this paper, namely the evasion of the obstacle described above and the robust structural
reasons for this evasion.

\section{Propagation \label{sec5}}

As shown in the last section and as anticipated by Smolin, repeated ultralocal actions of the finite triangulation
Hamiltonian  constraint on a kinematic state do not propagate quantum excitations over long ranges. 
In this section we reformulate the notion of propagation in terms of properties of physical states which lie in the
joint kernel of the diffeomorphism and Hamiltonian constraints.

Recall that solutions to the diffeomorphism constraint (see equation (\ref{diffavge})) 
do not lie in the kinematic Hilbert space. Instead they  are 
kinematically non- normalizable 
{\em distributions} which may be expressed as formal {\em sums}
over kinematic states. Hence a putative solution to both the Hamiltonian and 
the diffeomorphism constraints must also be  a distributional sum.
Consider such a solution and 
let a finest lattice charge net describing some discrete Cauchy slice with quantum matter be a summand 
in the sum which represents this solution. Recall, from section \ref{sec3} that this slice 
(together with its quantum matter) evolves under the action
of the finite gauge transformations $(\Phi_+, \Phi_-)$.
If the finest lattice charge net corresponding to {\em any}
finite  evolution of this discrete slice with quantum matter is also a summand in the sum which represents this
solution then we shall say 
that the solution encodes propagation effects.
\footnote{
As seen in section \ref{sec3} this statement of propagation holds for physical states obtained by group averaging.
So we could as well have formulated propagation as the condition  that the kernel of the diffeomorphism and 
Hamiltonian constraints be identical to the physical state space obtained by group averaging.
We choose to formulate the statement in the way we have done because such a formulation generalises more easily to 
the case of LQG where we do not have the possibility of group averaging over the transformations generated by 
{\em all} the constraints .
}

In the previous section we isolated the key obstruction to long range evolution by repeated actions of the Hamiltonian
constraint. We now show how our new  formulation of propagation overcomes this obstruction.
Specifically, as in section \ref{sec4}, let us consider 3 successive edges  of a finest lattice charge net 
with  embedding charge 
sequence $(k_3^+,k_3^-),(k_1^+,k_1^-),(k_2^+,k_2^-)$ with succesive vertices $u$ and $v$ so that 
$u$ is the `3-1' vertex and  $v$ is the `1-2' vertex. Recall from section \ref{sec4} that repeated actions of the 
finite triangulation 
Hamiltonian constraint on this charge net are unable to produce  the  charge net with embedding charge sequence 
$(k_3^+,k_3^-), (k_3^+,k_1^-),(k_1^+,k_2^-), (k_2^+,k_2^-)$ (see equation (\ref{no1})), this inability 
being the key obstruction to the generation of long range evolution through such repeated actions.
We now show that if a physical state in the  kernel of the diffeomorphism and Hamiltonian constraints
has the original `$(k_3^+,k_3^-),(k_1^+,k_1^-),(k_2^+,k_2^-)$' chargenet as a summand, it {\em necessarily}
has a chargenet with the desired $(k_3^+,k_3^-), (k_3^+,k_1^-),(k_1^+,k_2^-), (k_2^+,k_2^-)$ sequence.
We proceed as follows.

Let $\Psi$ be a distribution represented by a sum of `bra' states. We shall say that 
a chargenet 
$|\gamma, \vec{k^+}, \vec{k^-}, \vec{l^+}, \vec{l^-} \ket$ is  in $\Psi$ iff the bra
$\bra\gamma, \vec{k^+}, \vec{k^-}, \vec{l^+}, \vec{l^-}|$ is a summand in the sum over bras which 
represents $\Psi$. Next recall that $\Psi$ is a solution to the continuum limit of the finite 
triangulation Hamiltonian constraint (\ref{chat}) iff
for every $|\gamma, \vec{k^+}, \vec{k^-}, \vec{l^+}, \vec{l^-} \ket$ we have that 
\be
\lim_{\delta \rightarrow 0}\Psi (-i\hbar \sum_v N(v)\frac{ {\hat U}_{ham, \delta , v} -{\bf 1} }{\delta }
|\gamma, \vec{k^+}, \vec{k^-}, \vec{l^+}, \vec{l^-} \ket) =0 .
\label{solh}
\ee
Now suppose that $\Psi$ is a solution to 
the Hamiltonian constraint
(\ref{solh}) and that 
$|\gamma, \vec{k^+}, \vec{k^-}, \vec{l^+}, \vec{l^-} \ket$ is in  $\Psi$. Then it must be the case that for 
all sufficiently small $\delta$ that ${\hat U}_{ham, \delta , v}|\gamma, \vec{k^+}, \vec{k^-}, \vec{l^+}, \vec{l^-} \ket$
is also in $\Psi$ (else the nontrivial  contribution from  the `${\bf 1}$' term in (\ref{solh})
will not be cancelled). 
A similar arguementation leads to the converse namely that if
$\Psi$ 
satisfies (\ref{solh}) and if 
${\hat U}_{ham, \delta , v}|\gamma, \vec{k^+}, \vec{k^-}, \vec{l^+}, \vec{l^-} \ket$ is in $\Psi$
for all sufficiently small $\delta$,
then $|\gamma, \vec{k^+}, \vec{k^-}, \vec{l^+}, \vec{l^-} \ket$ is also in $\Psi$.

Similarly, $\Psi$ is a solution to the continuum limit of the finite triangulation 
Hamiltonian constraint(\ref{chata}) iff
for every $|\gamma, \vec{k^+}, \vec{k^-}, \vec{l^+}, \vec{l^-} \ket$ we have that 
\be
\lim_{\delta \rightarrow 0}\Psi (i\hbar \sum_v N(v)\frac{ {\hat U}^{\dagger}_{ham, \delta, v} -{\bf 1} }{\delta }
|\gamma, \vec{k^+}, \vec{k^-}, \vec{l^+}, \vec{l^-} \ket) =0 
\label{solha}
\ee
Similar arguments imply that for $\Psi$ 
which is a solution to 
the Hamiltonian constraint
(\ref{solha}),  iff $|\gamma, \vec{k^+}, \vec{k^-}, \vec{l^+}, \vec{l^-} \ket$ is in  $\Psi$ then for all 
for sufficiently small $\delta$, 
${\hat U}^{\dagger}_{ham, \delta, v}|\gamma, \vec{k^+}, \vec{k^-}, \vec{l^+}, \vec{l^-} \ket$  is in  $\Psi$.

It is also useful for what follows to recall the following: \\
(a) the action of ${\hat U}_{ham, \delta , v}$ on any charge net 
$|\gamma, \vec{k^+}, \vec{k^-}, \vec{l^+}, \vec{l^-} \ket$
is such that for small enough $\delta >0$ the chargenets 
${\hat U}_{ham, \delta , v}|\gamma, \vec{k^+}, \vec{k^-}, \vec{l^+}, \vec{l^-} \ket$ are in the {\em same}
diffeomorphism class,  \\
(b) a similar statement holds for  ${\hat U}_{ham, \delta , v}^{\dagger}$,\\
(c) $\Psi$ is a solution to the diffeomorphism constraints iff it is a linear combination of states
of the form (\ref{diffavge}).

From the above discussion, it is straightforward to check that  the following Lemma {\bf L1} holds.
\\

\noindent {\bf L1}: Let  $\Psi$ be a solution to the diffeomorphism constraint, the Hamiltonian 
constraint in (\ref{solh}) and the Hamiltonian constraint in (\ref{solha}). Then the following  statements hold:\\

(i)
The charge net $|\gamma, \vec{k^+}, \vec{k^-}, \vec{l^+}, \vec{l^-} \ket$ is in  $\Psi$ iff all diffeomorphic images of $|\gamma, \vec{k^+}, \vec{k^-}, \vec{l^+}, \vec{l^-} \ket$ are in 
$\Psi$,\\

(ii)The charge net $|\gamma, \vec{k^+}, \vec{k^-}, \vec{l^+}, \vec{l^-} \ket$ is in  $\Psi$ iff  
for all sufficiently small $\delta >0$, 
${\hat U}_{ham, \delta , v}|\gamma, \vec{k^+}, \vec{k^-}, \vec{l^+}, \vec{l^-} \ket$ is in $\Psi$,
\\

(iii) The charge net $|\gamma, \vec{k^+}, \vec{k^-}, \vec{l^+}, \vec{l^-} \ket$ is in  $\Psi$ iff for all sufficiently small $\delta >0$, 
${\hat U}^{\dagger}_{ham, \delta , v}|\gamma, \vec{k^+}, \vec{k^-}, \vec{l^+}, \vec{l^-} \ket$ is in $\Psi$,
\\

\noindent where in  (ii) (respectively (iii)) `sufficiently small' means `sufficiently small that the diffeomorphism class 
of ${\hat U}_{ham, \delta , v}|\gamma, \vec{k^+}, \vec{k^-}, \vec{l^+}, \vec{l^-} \ket$ (respectively
${\hat U}^{\dagger}_{ham, \delta , v}|\gamma, \vec{k^+}, \vec{k^-}, \vec{l^+}, \vec{l^-} \ket$) remains the same
for all such $\delta$.
\footnote{
This characterization of `sufficiently small'  follows from (a)-(c) 
and the fact that $\Psi$ in {\bf L1} is, in particular, a solution to the 
 diffeomorphism constraint.}
\\

We are now ready to state and prove our desired result.
\\

\noindent{\em Proposition}: Let $\Psi$ be a solution to the diffeomorphism constraint, the Hamiltonian
constraint (\ref{solh}) and the Hamiltonian constraint (\ref{solha}). 
Let $|\gamma, \vec{k^+}, \vec{k^-}, \vec{l^+}, \vec{l^-} \ket$ be
 a finest lattice chargenet in  $\Psi$. Let $|\gamma, \vec{k^+}, \vec{k^-}, \vec{l^+}, \vec{l^-} \ket$ 
have  3 consecutive edges
\footnote{ The proof below applies independent whether the 2 vertices associated with these edges are 
spacelike or not.}
 with embedding charges
$(k_3^+,k_3^-),(k_1^+,k_1^-),(k_2^+,k_2^-)$.
Then the chargenet with these 3 edges replaced by 4 successive
edges with the  embedding charge sequence $(k_3^+,k_3^-), (k_3^+,k_1^-),(k_1^+,k_2^-), (k_2^+,k_2^-)$ 
is {\em necessarily} in $\Psi$. 
\\

\noindent{\em Proof:}  Denote the `1-2' vertex of $|\gamma, \vec{k^+}, \vec{k^-}, \vec{l^+}, \vec{l^-} \ket$
by $u$ and the `2-3'  vertex  by $v$. There are 3 steps to the proof:\\
\noindent{\em Step 1}: 
Act with ${\hat U}_{ham, \delta_1 , v}$ on $|\gamma, \vec{k^+}, \vec{k^-}, \vec{l^+}, \vec{l^-} \ket$
at its `1-2' vertex $v$.
We obtain the chargenet with the above sequence replaced by 
\be
(k_3^+,k_3^-),(k_1^+,k_1^-),(k_1^+, k_2^-), (k_2^+,k_2^-). 
\label{1}
\ee
{\bf L1}(ii) implies the chargenet so obtained is also in $\Psi$ (we have implicitly chosen
$\delta_1$ small enough that {\bf L1}(ii) applies).

\noindent{\em Step 2:} Act on the chargenet obtained at the end of Step 1  with 
${\hat U}_{ham, \delta_2 , u}$  on its `3-1' vertex $u$ to get the sequence 
\be
(k_3^+,k_3^-),(k_3^+, k_1^-), (k_1^+,k_1^-),(k_1^+, k_2^-), (k_2^+,k_2^-)
\label{2}
\ee
{\bf L1} (ii) implies the chargenet so obtained is also in $\Psi$ (similar to Step 1, we have implicitly chosen
$\delta_2$ small enough that {\bf L1}(ii) applies).

\noindent{\em Step 3:} Consider the (desired) chargenet with sequence 
\be
(k_3^+,k_3^-), (k_3^+,k_1^-),(k_1^+,k_2^-), (k_2^+,k_2^-)
\label{seqd}
\ee
and call the vertex at the intersection of the $(k_3^+,k_1^-)$ and $(k_1^+,k_2^-)$ edges as $w$.
Act on this chargenet by ${\hat U}^{\dagger}_{ham, \delta , w}$ for sufficiently small $\delta$ in the sense
of (b) above. It follows that for every such $\delta >0$, this action yields 
a chargenet with the sequence (\ref{2}) which is diffeomorphic to the particular chargenet obtained at the 
end of Step 2. 
Since the latter chargenet is in $\Psi$, {\bf L1}(i) implies that the chargenets, obtained by the action of 
 ${\hat U}^{\dagger}_{ham, \delta , w}$ for all sufficiently small $\delta$ on the desired chargenet 
with sequence (\ref{seqd}), are in $\Psi$. The (converse of) {\bf L1}(iii) then implies that the 
 desired chargenet is also in $\Psi$.
This completes the proof.
\\

Note that the  first two steps of the proof involve actions by the Hamiltonian constraint
on the chargenet in question. Hence, from section \ref{sec4} these steps by themselves are incapable of generating 
the desired result. It is  Step 3 which is the {\em key} step. 
The success of this step hinges on the imposition, as a constraint, 
of the `kinematic adjoint' (\ref{chata}),(\ref{solha}) of the constraint (\ref{chat}),(\ref{solh}) 
together with its `${\hat U}^{\dagger} -1$' structure.

Finally, we also note that, acting by 
${\hat U}_{ham, \delta , w}$ on the  
chargenet  (\ref{seqd}) 
with sequence
\be
(k_3^+,k_3^-), (k_3^+,k_1^-),(k_1^+,k_2^-), (k_2^+,k_2^-)
\ee
at its vertex $w$ (where the $(k_3^+,k_1^-)$ and $(k_1^+,k_2^-)$ edges meet)  we obtain a chargenet with sequence
\be
(k_3^+,k_3^-), (k_3^+,k_1^-),(k_3^+,k_2^-),(k_1^+,k_2^-), (k_2^+,k_2^-). 
\label{2disp}
\ee
If in the original chargnet the vertices 
$u,v$ are spacelike, it is easy to see that the point 
$(\hbar k_3^+, \hbar k_2^-)$ represents a 2 lattice displacement from the original set of points corresponding to 
$(\hbar k_3^+,\hbar k_3^-),(\hbar k_1^+,\hbar k_1^-),(\hbar k_2^+,\hbar k_2^-)$. Thus as indicated in section \ref{sec1}
we are able to demonstrate evolution beyond 1 lattice displacement to 2 lattice displacements.
As mentioned at the end of section \ref{sec4}, the demonstration of long range evolution is relegated to the Appendix.

\section{Discussion \label{sdiscuss}}

Before we proceed to more general remarks, we comment on the derivation of the constraint action 
(\ref{chata}) from choices of finite triangulation approximants to the local fields 
which comprise the Hamiltonian constraint. The reader not interested in fine technical details of 
our prior work \cite{polypft3} may skip the next two paragraphs and go on to peruse the more general remarks.

First consider the derivation of the action (\ref{chat}). In this regard
Reference \cite{polypft3}
provides a  detailed derivation of finite triangulation approximants to $H_+, H_-$ and it is from these
that the approximant to the Hamiltonian constraint operator (equation (112),\cite{polypft3}) is obtained. 
It is straightforward
to see that this  is the same as (\ref{chat}) albeit in a slightly different notation.
Recall that in order to obtained the desired constraint operator action (112),\cite{polypft3},
 the embedding sector approximants to the local fields in the constraint  
are constructed straightforwardly in \cite{polypft3},  
first as appropriate
classical approximants involving state dependent charges, and then as operators. Also recall that Reference 
\cite{polypft3}
is unable 
derive   matter 
sector approximants in this way. More in detail, that work is
unable to construct classical approximants to the 
local matter fields in the Hamiltonian constraint such that their replacement by operators is
consistent with the desired action (112) of \cite{polypft3}
\footnote{We think that the underlying reason for this inability is that, as mentioned in 
\ref{sqtk}, there is no `unpolymerised' matter variable.}. Hence an indirect appeal to the Hamiltonian 
vector fields of the matter part of the constraints is made \cite{polypft3} and this constitutes a slight departure
from the strict `Thiemann-like' prescription.

Next consider the derivation of the adjoint action (\ref{chata}). It turns out that
the action (\ref{chata}) requires a slightly different choice of  approximants to local fields
in the constraints. It is once again  straighforward to construct the desired embedding momenta approximants
\footnote{See the discussion immediately after (47),\cite{polypft3}. There is a typographical error
in the choice of approximant for the `+' embedding momenta which is claimed to lead to the leftward
displacement of the `+' vertex: the subscript on the embedding holonomy should be $\triangle -1$
instead of $\triangle$.}. However, for the matter sector one needs to again consider Hamiltonian vector fields.
While we have not done this in detail, we anticipate that the considerations of Section VB,\cite{polypft3}
may be mimicked with slightly different choices so as to obtain an action which contributes appropriately
to the Hamiltonian constraint approximant so as to obtain (\ref{chata}). More generally, our current viewpoint
is that it is the Hamiltonian vector field structure of the constraint rather than the constraint itself 
which is primary and that this Hamiltonian vector field structure is what one should import
in a suitable fashion into  quantum theory even if one is unable to provide concrete classical
approximants to the constraint itself. From this viewpoint one can directly posit the actions (\ref{chat}), (\ref{chata})
as approximants to the infinitesmal transformations generated by the constraint (\ref{cham}) 
without unduly worrying about classical approximants to the constraint itself.

Finally, in this work we have {\em not} imposed the 
zero mode constraint \cite{polypft1,polypft2} as we feel that it does not impinge on the issues
we are concerned with here. For completeness, this constraint should be imposed (else the classical arena is not a 
a  phase space \cite{polypft1,polypft2}). Since this constraint commutes
with the action of finite gauge transformations, it commutes with the action of finite spatial diffeomorphisms,
 and the action of the finite triangulation Hamiltonian constraint (\ref{chat}), (\ref{chata}). 
Since it commutes with the latter at any value of the  triangulation parameter $\delta$, we expect it to commute
with the continuum limit action of (\ref{chat}),(\ref{chata}). Hence  we expect that it shouldnt matter whether
we find the kernel of the Hamiltonian and diffeomorphism constraints first and then group average these solutions
over the zero mode constraint or whether we first solve the zero mode constraint by group averaging, define the 
Hamiltonian and diffeomorphism constraints on the resulting states and then find the kernel of the Hamiltonian 
and diffeomorphism constraints. Verifying the above expectations, while straightforward,
 lies outside the scope of this work. Incidentally, we believe that in hindsight, the treatment of the zero
mode in \cite{polypft3} was too perfunctory, that the arguements there should be seen as arguements prior to the 
implementation of the zero mode constraint and that a proper treatment of the zero mode constraint 
should explicitly verify our expectations as stated above.
This concludes our discussion of fine technical matters in polymer PFT.

We now discuss the key structural lessons from this work for LQG. 
Let us refer to the new charge nets  obtained by the action of a constraint  on 
a given chargenet as `children' of this `parent' chargenet. 
In this language Smolin's general considerations imply, correctly, that such  children and their descendants do not 
encode long range propagation. However, given the structure of the constraints (\ref{chat})- (\ref{chata}),
Lemma {\bf L1} implies that if 
a parent is in $\Psi$ so are its children, and, conversely  if any child is in $\Psi$ then so are 
all its parents. It is the converse statement which provides the key ingredient of `non-unique parentage' in 
the crucial Step 3 of our proof in the previous section.

From a general point of view what Step 3 effectively achieves is the merging of two vertices of a `child'  into 
a single vertex of a `parent' (the single vertex being $w$ in our proof). Note that this merging cannot  be 
achieved by the action (\ref{chat}) nor by its kinematic adjoint action (\ref{chata}) on the child.
This is apparent from the considerations of section \ref{sec3} which apply equally to both actions, the 
key point being that these actions are only defined as finite triangulation constraint actions  
{\em for sufficiently small} $\delta$. It is this caveat of ``{\em for sufficiently small} $\delta$''
that leads to ultralocality and the impossibility of merging vertices of the child by action on the child.
Rather, the merging is achieved by {\em seperating} vertices of the parent via the action of the kinematic
adjoint and using the structure of the constraint action in terms of the difference of a unitary operator
and the identity to conclude that the existence of children in a physical state impy the existence 
of all their possible ancestors. Once the 2 vertices of the child have been effectively merged through this
structure to yield the desired parent, a suitably chosen  action of the Hamiltonian constraint (see the last paragraph of 
\ref{sec5})  on this
parent creates a {\em different} child and the sequence `parent $\rightarrow$ child $\rightarrow$ different
parent $\rightarrow$ different child' constitutes an evolution path to a final discrete Cauchy slice
two lattice spacings away. This sequence may be viewed as the propagation of a perturbation
(namely the $k_2^-$ charge together with the associated `-' matter charges) `leftward' along the charge net.

From the above discussion it is apparent that the key structures responsible for propagation are exactly (i) and (ii)
of section \ref{sec1} and that propagation should be viewed as encoded in the structure of physical states rather
than
as a property of repeated actions of the finite triangulation Hamiltonian constraint on kinematical states.
To conclude, while we do expect the general Thiemann procedure to yield a Hamiltonian constraint
with ultralocal action, 
we are optmistic that the structural lessons arising out of  this work can be imported in a suitable way to LQG 
so as to restrict the choice of this ultralocal action  in such a way that 
physical states in the kernel of the Hamiltonian (and diffeomorphism) constraints do encode propagation effects.
\footnote{See Footnote \ref{fn3} in this regard.}

\section*{Acknowledgements} I thank Lee Smolin for raising the general issues dealt with in this work
during the Loops 13 conference. This work was motivated by a conversation with Michael Reisenberger
during that conference and I am very grateful for his remarks in the course of that 
conversation.

\appendix
\section{Proof of long range propagation}
In this appendix we assume familiarity with the contents of \cite{polypft1,polypft2,polypft3}.
We shall also set $\hbar=1$ by a suitable choice of units.

Recall the following from \cite{polypft2}. The diffeomorphisms of the circle,
 $\Phi_{\pm}$, admit periodic extensions to the 
real line, also denoted by $\Phi_{\pm}$. As in \cite{polypft2}, denote the set
$(\gamma_{\pm}, \vec{k^{\pm}}, \vec{l^{\pm}})$ by $s^{\pm}$  and the corresponding states
$|\gamma_{\pm}, \vec{k^{\pm}}, \vec{l^{\pm}}\ket $ by $|s^{\pm}\ket$
with $|s^+\ket \otimes |s^-\ket := |s^+, s^-\ket$. 
$s^{\pm}$ is referred to as a charge network or charge network label and $|s^{\pm}\ket$ as a charge network state. 
Whereas $s^{\pm}$ is defined on $[0,2\pi]$ its extension $s^{\pm}_{ext}$ is defined on the entire real line
by periodic extension of the graph $\gamma^{\pm}$, its edges and its matter charge labels and quasi-periodic
extension (with appropriate augmentation by factors of $\pm L$) of its embedding charge labels. 
The periodic extensions of $\Phi_{\pm}$ have a well defined action on $s^{\pm}_{ext}$.
The state defined by the  restriction of the image of this action to the interval $[0,2\pi]$ coincides
with that obtained by the action of ${\hat U}_{\pm}(\Phi_{\pm})$ on $|s^{\pm}\ket$  \cite{polypft1,polypft2} so that :
\be
{\hat U}_{\pm}(\Phi_{\pm})|s^{\pm}, \ket = |s^{\pm}_{\Phi_{\pm}}     \ket, 
\;\;\;\; s^{\pm}_{\Phi_{\pm}} := \Phi_{\pm}(s^{\pm}_{ext})|_{[0,2\pi]} .
\ee
Clearly, the  action of $\Phi_{\pm}$ on {\em any} interval of coordinate length $2\pi$ determines
its action everywhere on the real line by virtue of the periodicity of this action. Similarly the restriction 
of $s^{\pm}_{ext}$ to any interval of coordinate length $2\pi$ determines 
$s^{\pm}_{ext}$ everywhere on the real line.

Since ${\hat U}_{ham, \delta, v}, {\hat U}^{\dagger}_{ham, \delta, v}$
are constructed from ${\hat U}_{\pm}(\phi_{\delta, v}), {\hat U}^{\dagger}_{\pm}(\phi_{\delta, v})$ it follows that 
the action of the finite triangulation Hamiltonian constraint on any state $|s^+,s^-\ket$ 
is determined by the action of the 
diffeomorphisms $\Phi_{\pm}=  \phi_{\delta, v}$  (and their inverses) on the restriction of $s^{\pm}_{ext}$ 
to any interval of
coordinate length $2\pi$. 


More in detail,  for some real $y$, consider the interval $[y, y+2\pi]$. $\Phi_{\pm}$ maps this interval to 
the interval $[y^{\pm}, y^{\pm}+ 2\pi]$ where we have set $\Phi_{\pm}(y) =: y^{\pm}$.
Consider the restriction,  $s^{\pm}_{ext}|_{[y, y+2\pi]}$  of $s^{\pm}_{ext}$ to the interval
$[y, y+2\pi]$. Clearly $\Phi_{\pm}$ has a natural action on $s^{\pm}_{ext}|_{[y, y+2\pi]}$  (it maps
every edge $e$ of the graph underlying  $s^{\pm}_{ext}|_{[y, y+2\pi]}$ into its image $\Phi_{\pm}(e)$ in 
$[y^{\pm}, y^{\pm}+ 2\pi]$, with $\Phi_{\pm}(e)$ being colored by the same charges as $e$). 
Denote the resulting charge net on $[y^{\pm}, y^{\pm}+ 2\pi]$ by 
$\Phi_{\pm}(s^{\pm}_{ext}|_{[y, y+2\pi]})$.
It is straightforward to check that  $\Phi_{\pm}(s^{\pm}_{ext}|_{[y, y+2\pi]})$
is just the restriction of $\Phi_{\pm}(s_{ext})$ to the interval  $[y^{\pm}, y^{\pm}+ 2\pi]$.
It then follows that the extension, $(\Phi_{\pm}(s^{\pm}_{ext}|_{[y, y+2\pi]}))_{ext}$,
 of the charge net $\Phi_{\pm}(s^{\pm}_{ext}|_{[y, y+2\pi]})$
to the real line is just  $\Phi_{\pm}(s_{ext})$. Finally, the restriction of this extension to $[0,2\pi]$\
is the just the charge net $s^{\pm}_{\Phi_{\pm}}$ i.e.
\be
(\Phi_{\pm} (s^{\pm}_{ext}|_{[y, y+2\pi]})_{ext}   )|_{[0,2\pi]}
= s^{\pm}_{\Phi_{\pm}} .
\ee
Since ${\hat U}_{\pm}(\Phi_{\pm})|s^{\pm}\ket = |s^{\pm}_{\Phi_{\pm}}\ket$, 
the content of  this paragraph  is just a transcription to mathematical notation of what 
we said in words in the previous paragraph.

The above discussion implies 
that {\bf L1}, Section \ref{sec5} may be rephrased 
as follows in the notation used above (this rephrasing, while cumbersome and seemingly roundabout, is useful for
our purposes in this appendix).
\\

\noindent {\bf L2}: Let  $\Psi$ be a solution to the diffeomorphism constraint, the Hamiltonian 
constraint in (\ref{solh}) and the Hamiltonian constraint in (\ref{solha}). Let 
$|s^+, s^-\ket$ be a finest lattice
chargenet state.
The charge net state $|s^+, s^-\ket$ is in  $\Psi$ iff the set of charge net states $\{|t^+,t^-\ket\}$ is 
in $\Psi$ where the elements $|t^+,t^- \ket$ of this set are defined by any of  (i)- (iii) below, 
with $y$ any fixed real number
and $v \in [y, y+2\pi ]$:\\

(i) $(\phi (t^{\pm}_{ext})|_{[y, y+2\pi]})_{ext}|_{[0,2\pi]} = s^{\pm}$ for any diffeomorphism $\phi$. \\

(ii) $t^+ = (\phi_{\delta, v} (s^+_{ext}|_{[y, y+2\pi]}))_{ext}|_{[0,2\pi]}$,  
$\;\;t^- = ( \phi^{-1}_{\delta, v} (s^-_{ext}|_{[y, y+2\pi]}) )_{ext}|_{[0,2\pi]}$ for all sufficiently small $\delta$.
\\

(iii) $t^+ = ( \phi^{-1}_{\delta, v} (s^+_{ext}|_{[y, y+2\pi]}) )_{ext}|_{[0,2\pi]}$,  
$\;\;t^- =  (  \phi^{}_{\delta, v} (s^-_{ext}|_{[y, y+2\pi}]))_{ext}|_{[0,2\pi]}$ for all sufficiently small $\delta$.
\\

\noindent where in  (ii)-(iii) `sufficiently small' means `sufficiently small that the 
diffeomorphism class of  $|t^+, t^-\ket$ does not change.
\footnote{Recall that $\phi_{\delta, v}$ is such that $\phi_{\delta, v}$ moves $v$ to the right by a coordinate
distance $\delta$, its inverse $\phi^{-1}_{\delta, v}$ moves $v$ to the left by a distance $\delta$ and $\phi_{\delta, v}$
is identity outside an interval of size of  order $\delta$ about $v$.}
\\

Next note that similar to (\ref{ptn}) we may consider a fine enough graph which underlies both $s^+$ and $s^-$
and whose edges carry both $+$ and $-$ charges. Let us denote the resulting label set
$(\gamma, \vec{k^+}, \vec{k^-}, \vec{l^+}, \vec{l^-})$ by $s$ and set 
\be 
|s\ket:= |s^+, s^-\ket .
\ee
We may also define the extended
label $s_{ext}$ by a periodic extension $\gamma_{ext}$ of $\gamma$, a periodic extension of the matter charge labels 
 and appropriate (quasi)periodic
extensions of the embedding charges. Clearly, $\gamma_{ext}$ constitutes  a fine enough 
graph which underlies both $s^+_{ext}$ and $s^-_{ext}$, and $s_{ext}$ accomodates both the $+$ and $-$ charge labels
of $s^+_{ext}$ and $s^-_{ext}$.

The Proposition of Section \ref{sec5} can then be rephrased as:
\\

\noindent{\bf P1}: Let $\Psi$ be a solution to the diffeomorphism constraint, the Hamiltonian
constraint (\ref{solh}) and the Hamiltonian constraint (\ref{solha}). 
Let $|s\ket= |s^+, s^-\ket$ be
 a finest lattice chargenet in  $\Psi$. Let $y$ be some real number. Let the restriction
of $s_{ext}$  to the interval $[y, y+2\pi]$ be 
$s^{}_{ext}|_{[y, y+2\pi]}$.
Let the graph underlying $s^{}_{ext}|_{[y, y+2\pi]}$
have  3 consecutive edges $e_3, e_1, e_2$ with 
$e_3,e_1,e_2 \subseteq [y, y+2\pi]$
with embedding charges
$(k_3^+,k_3^-),(k_1^+,k_1^-),(k_2^+,k_2^-)$.
Consider the charge net  state  $|s'\ket =|s'^+, s'^-\ket$ such that  
$s'_{ext}|_{[y, y+2\pi]}$ agrees with $s_{ext}|_{[y, y+2\pi]}$
except that these 3 edges  of the latter are  replaced in the former by 4 successive
edges $e'_4, e'_{3}, e'_1,e'_2$ with the  embedding charge sequence 
$(k_3^+,k_3^-), (k_3^+,k_1^-),(k_1^+,k_2^-), (k_2^+,k_2^-)$.
Then $|s' \ket$ is {\em necessarily} in $\Psi$. 

It is straightforward to check that {\bf P1} can be proved along the lines of the proof of the Proposition
of section \ref{sec5}
\\

Next, let $s,y,s', \Psi$ be as in {\bf P1}. Clearly,
there exists a diffeomorphism $\phi$ which is identity to the right of $e_2$ and the left of $e_3$ in $[y,y+2\pi]$
 and which maps $s'$ to some $s''$ such that in 
$s''_{ext}|_{[y, y+2\pi]}$ we have that $\phi (e'_2)= e_2$, $\phi (e'_1) = e_1$, $\phi (e'_4\cup e'_3)= e_3$. Thus
$s''_{ext}|_{[y, y+2\pi]}$ agrees with $s_{ext}|_{[y, y+2\pi]}$ outside the interval $\phi (e'_3) \cup e_1$.
{\bf L2} (i) then implies that the following corollary to {\bf P1} holds:\\

\noindent{\bf C1}: 
Given $s,y,s',s''$ as above, $|s''\ket$ is also in $\Psi$.
\\

With these results in place we now show that if a finest lattice charge net state $|s\ket$  is 
in any diffeomorphism invariant solution $\Psi$ to the 
Hamiltonian constraints (\ref{solh}) and (\ref{solha}), then all states related to 
$|s\ket$  by the action of finite gauge transformations generated by $H_+,H_-$
are also 
in $\Psi$. It then follows from the notion of propagation introduced in section \ref{sec5}  that $\Psi$ 
encodes long range propagation effects. We proceed by proving the following lemmas {\bf L3}- {\bf L7}.

In the proof of {\bf L3} below we shall use the following notion of `{\em sequence of `-' embedding  charges}'. 
Let $s= (s^+,s^-)$ be a finest lattice chargenet. Let $s^-_{ext}$ be the extension of $s^-$
and consider the restriction $s^-_{ext}|_{[y, y+2\pi]}$ to some $2\pi$ interval $[y, y+2\pi]$.
Consider any fine enough graph underlying 
the charges on $s^-_{ext}|_{[y, y+2\pi]}$. Let the edges of this graph be $e_I,I=1,..,B$ where 
$e_J$ is to the right of $e_I$ for $J>I$. Let the `-' embedding charge on $e_I$ be $k^-_I$. Then the ordered set of `-' charges $(k^-_1, k^-_2, .., k^-_B)$ 
is referred to as the sequence of `-' embedding charges associated with  $s_{ext}|_{[y, y+2\pi]}$.
The finest lattice property implies that this sequence is non-increasing and that for the coarsest graph
underlying $s^-_{ext}|_{[y, y+2\pi]}$ this sequence is strictly decreasing. Thus, 
depending on the fineness of the graph,  there may be several instances of  a number of   successive entries 
in the sequence being  identical. 
In the proof of {\bf L3} below we shall refer to such sequences directly without explicitly constructing the 
graphs which define them; however it is to be understood that such graphs exist (as the reader may verify, 
it is straightforward to show their  existence).
\\


\noindent{\bf L3} Let $|s\ket =|s^+,s^-\ket$ be a finest lattice state which is 
in a solution $\Psi$ to the 
diffeomorphism constraint and the Hamiltonian constraints (\ref{solh}) and (\ref{solha}).
Let $s'^+=s^+$. Let $s'^-$ be such that the set of its matter charge labels, ordered edgewise from left to right 
is identical to the corresponding set for $s^-$ and such that the set of its  embedding charge labels,
ordered from left to right are obtained by decreasing each of the elements of the corresponding set for $s^-$ by
 $M{L}$ for some arbitrary positive integer $M$.
Then $|s'\ket = |s'^+, s'^-\ket$ is also in $\Psi$.
\\

\noindent{\em Proof}: As in Footnote \ref{fnN}, we shall assume  $N>>4$.
Let the coarsest graph underlying $s$ have  edges $e_I, I=1,..,A$ with embedding charges $(k^+_I, k^-_I)$ with 
$e_I$ to the left of $e_J$ for $J>I$. Clearly $A>>4$. Also
note that the chargenets in {\bf L2}, {\bf P1}, {\bf C1} are related by the action of
gauge transformations so that these chargenets are all in the finest lattice sector and respect properties (i)- (iii),
section \ref{sec3}. We shall implicitly use this fact repeatedly in what follows.

 Consider $s_{ext}$ and let $e_0$ be the edge (of the coarsest graph underlying $s_{ext}$) 
in the interval $[-2\pi, 0]$ starting at some point $y_0 \in [-2\pi, 0]$ and
ending at the orgin. 
In the next 3 paragraphs we will repeatedly apply {\bf C1} to appropriately chosen triplets of edges in the interval 
$[y_0, y_0+2\pi]$, this
being the interval spanned by $e_I, I=0,..,A-1$. 

{\bf C1} applied to the 3 edges $e_{0}, e_1, e_2$ implies that we
can displace the `-' charge $k^-_1$ on the edge $e_1$ by extending the coloring $k_2^-$ to $e_1$ so that $e_1, e_2$
are colored by $k_2^-$. Denote the resulting extended charge net by $s_{1ext}$ so that its restriction  
to $[0,2\pi]$ is 
the charge net $s_1$. Clearly the  `-' embedding charge on the  last edge of $s_1$
(in the interval $[0,2\pi ]$)  is 
now $k_1^- -L$ (see \cite{polypft1,polypft2} and Footnote \ref{fn1}). 
Let  $e^{(1)}_0$ be the edge in $s_{1ext}$ which ends at $x=0$. The edge $e^{(1)}_0$
corresponds to the edge $\phi (e_3')$ and $e_0$ to $\phi (e_4'\cup e_3')$ in 
the remarks before {\bf C1}. It follows that  $e^{(1)}_0$
is contained in $e_0$. Hence, denoting the left end point of $e^{(1)}_0$ by $y{(1)}_0$, we have that 
$0> y^{(1)}_0 > y_0$.

Iterate this process to spread $k^-_3$ leftwards in $s_1$ at the cost of $k^-_2$ as follows. First note that 
the graph which underlies $s_1$ can be chosen such that its first 3 edges coincide with $e_1,e_2, e_3$.
From {\bf C1} we can spread $k_3^-$ to $e_2$. From {\bf P1} and {\bf C1} note that in the resulting chargenet
the edges
 $e_1, e_2, e_3$ can still be taken to be the first 3 edges with $e_1$ colored by $k_2^-$ and $e_2,e_3$ by $k^-_3$.
Next consider the extension of this chargenet and the edges $e^{(1)}_0, e_1,e_2$ therein.
Using {\bf C1} we can move $k^-_3$ from $e_2$ to $e_1$.
This yields the desired charge net $s_2$ with first edge colored by $k_3^-$ and last edge by $k^-_2-L$.
Denote the edge in $s_{2ext}$ ending at $x=0$ 
by $e^{(2)}_0$ and let its left end point be $y^{(2)}_0$. Here  $e^{(2)}_0$ corresponds to 
$\phi (e_3')$ and $e^{(1)}_0$ to $\phi (e'_4\cup e'_3)$  in the remarks before {\bf C1}. 
It follows that  {\bf C1} implies that $0 >y^{(2)}_0>y{(1)}_0 > y_0$.

Clearly, we can iterate this process such that after $A-2$ iterations we obtain $s_{A-2}$ with first charge 
$k^-_{A-1}$, last charge $k^-_{A-2} -L$  and the edge in $s_{A-2\;ext}$ 
ending at the origin with left endpoint $y_0^{(A-2)}$ such that $0 >y^{(A-2)}>
y^{(A-3)}...>y_0$.

Next,  consider the interval $[y_0^{(A-2)}, y_0^{(A-2)}+ 2\pi]$. 
In this paragraph we shall repeatedly apply {\bf C1} to appropriately chosen triplets of edges in this interval.
Note that the sequence of `-' charges in  $s_{A-2\;ext}$ restricted to this interval
 reads $(k^-_{A-2},k^-_{A-1},k^-_A, k^-_1-L,k^-_2-L,..,k^-_{A-3}-L)$. 
Repeated application of {\bf C1} to appropriate edges in this interval results in the spread of
$k^-_{A}$ leftwards  till $x=0$ resulting in the charge net $s_{A-1}$. The sequence of charges in 
$s_{A-1\;ext}$ restricted to this interval reads, as before, $(k^-_{A-2},k^-_{A-1},k^-_A, k^-_1-L,k^-2-L,..,k^-_{A-3}-L)$
except that now the edge ending at $x=0$ has charge $k^-_{A-1}$ and the edge starting at $x=0$ has charge $k^-_{A}$.
Finally, we repeatedly apply {\bf C1} to appropriate edges in this interval so as to spread $k^-_1-L$ leftwards
till the origin to yield the charge net $s_A$. The sequence of `-' charges in the restriction of 
$s_{A\;ext}$ to this interval is unaltered but now the edge starting out from the origin has charge $k^-_1-L$
and the edge ending at the origin has charge $k^-_A$.
This implies that the charge sequence in $s_A$ in the interval $[0,2\pi]$ reads \\
$(k^-_1-L,k^-_2-L,..,k^-_{A-3}-L, k^-_{A-2}-L, k^-_{A-1}-L,k^-_{A}-L)$.

Finally taking $s_A$ as the initial charge net and repeating the above procedure $M$ times leads to the 
desired result.\\

{\noindent}{\bf Note N1}: Interchanging the role of ${\hat U}_{\delta, v}$ and ${\hat U}^{\dagger}_{\delta,v}$ in 
the proof of  the Proposition of Section \ref{sec5} and in that of   {\bf P1}, {\bf C1} 
results in the rightward movement of the `-' charges. With this modification considerations 
identical to {\bf L3} lead to the `-' embedding charges being augmented by factors of $+ML$.
Similarly, it is straightforward to see that the Proposition of Section \ref{sec5} as well as {\bf P1},{\bf C1}
and {\bf L3} above can be modified to reflect leftward and rightward movement of the `+' charges with 
augmentation of the `+' charges by factors of $\pm ML$.
The fact that we are dealing with finest lattice charge nets satisfying (ii) of section \ref{sec3}
immeditiately implies that the matter charges are dragged along with the embedding charges yielding the 
desired charge configurations.
\\

\noindent{\bf L4} Let $|s\ket =|s^+,s^-\ket$ be a finest lattice state with $N^{\pm}$ 
edges as in  (iii) of Section \ref{sec3}. Let $|s\ket$ be in a solution $\Psi$ to the 
diffeomorphism constraint and the Hamiltonian constraints (\ref{solh}) and (\ref{solha}).
Then there exists $|s'\ket = |s'^+, s'^-\ket$ in $\Psi$ such that $s'^{\pm}$ have 
$N+1$ distinct embedding charges i.e. the coarsest graphs $\gamma'^{\pm}$ underlying $s'{\pm}$ have
$N^{\pm}= N+1$ edges.
\\

\noindent{\em Proof}: The proof consist of the following 2 steps.

{\em Step 1}: If $N^-=N+1$ proceed to Step 2. If not then we have  $N^- =N$. 
From (ii), (iii) Section \ref{sec3}, this  means that  
the embedding charge sequence on $s^-$
is of the form $k^-_{1}, k^-_2,...,k^-_N$ with $k^-_1 - k^-_N =L -a$.  Application of ${\hat U}_{\delta, v=0}$,
for small enough $\delta$, 
to $|s\ket$
drags part of the first edge of $s^-$ leftwards. The rightmost charge of resulting  chargenet $s_1$ is then,
by (quasi)periodicity, $k^-_1- L$ and the left most charge is still $k^-_1$ so that now $N^-=N+1$.
By {\bf L2}, the resulting  chargenet state $|s_1\ket$ is in $\Psi$.

{\em Step 2}:
In the charge net state $|s_1\ket$ at the end of Step 1, if $N^+=N+1$ we are done. If $N^+= N$ , an  application of 
${\hat U}_{\delta^{\prime}, v=0}$ 
for small enough $\delta^{\prime}$ to $|s_1\ket$  leads to the  
rightward dragging of part of the first edge of $s_1^+$ while
maintaining $N^-=N+1$.
It is then easy to see that the resulting charge net has $N^+= N+1$. Further the resulting charge net is in
$\Psi$ by {\bf L2}.

This completes the proof.
\\

\noindent {\bf L5} Let $|s\ket, |s'\ket$ be finest lattice charge nets with $|s'\ket$ in   $\Psi'$  where
$\Psi'$ is a solution 
to the diffeomorphism
constraint and the Hamiltonian constraints (\ref{solh}), (\ref{solha}). Let $s^{\pm},s'^{\pm}$ have $N+1$
distinct `$\pm$' embedding charges. Then there exist $|t'\ket$ in  $\Psi'$ such that
$s^{\pm}, t'^{\pm}$ have identical sets of ${\pm}$ embedding charge labels.
\\

\noindent{\em Proof}: Let us number the edges (and their embedding charge labels) 
in $s^-,s'^-$ from 1,..N+1 as we proceed rightwards 
 on the coarsest graphs underlying these charge nets. Thus  the left most charge in $s^-$ is $k^-_1$. 
By the finest lattice property  there exists a unique charge label $k'^-_m$ in $s'^-$  with the property that 
$k'^-_m$ is the largest 
\footnote{This rules out $m=N+1$}
`-' embedding charge in $s'^-$ such that  
$k'^-_m -k^-_1$ is an integer multiple `$M_-$' times $L$ .  If $m=1$ then we can apply {\bf L3} to construct 
$t^-$ such that its `-' embedding charge set agrees with that of $s^-$.

If $m\neq 1$, we apply {\bf C1} repeatedly to $s'$ so as to 
spread $k'^-_m$ leftwards
in a manner identical to that employed in the proof of {\bf L3}. As a result, on the resulting chargenet 
$k'^-_m$ becomes the first charge. Since $k'^-_m\neq k'^-_{m-1}$, it is easily verified that this charge net has 
$N$ distinct `-' charges. It is straightforward to verify that an application of  ${\hat U}_{\delta, v=0}$ for 
small enough $\delta$ moves the first
edge of the resulting chargenet slightly to the left so as to change $N^-$ to $N+1$ while maintaining the `+' embedding 
charge set.
Call the chargenet so obtained as $s'_1$. 

Next, we can identify the unique smallest `+' embedding charge  on $s'_1$ which differs from $k_1^+$ by an 
integer multiple
$M_+$ of $L$. If this charge colors the first edge of $s'_1$ we are done. If not,
we can spread this charge  
to the left of $s'$  by application of {\bf N1} so
that it becomes the first `+' embedding  charge on the resulting chargenet. 
An application of ${\hat U}_{ {\delta}^{\prime}, v=0 }$
for small enough $\delta^{\prime}$ on this chargenet with $N^+=N$  ensures that the resulting chargenet $s'_2$ 
has $N^+=N+1$
while maintaining $N^-=N+1$. 
As a result $s_2^{'\pm}$
has the same $\pm$ embedding chargesets as $s$ modulo factors of $M_{\pm} L$.

Finally we can apply {\bf L3}, {\bf N1} to obtain the desired charge net $t'$. From {\bf L1}, {\bf L3}, {\bf N1}
it follows that $|t'\ket$ is in $\Psi'$.
\\

\noindent {\bf L6} Let $|s\ket, |s'\ket$ be finest lattice charge nets with $|s'\ket$ in  $\Psi'$ where
$\Psi'$ is a solution 
to the diffeomorphism
constraint and the Hamiltonian constraints (\ref{solh}), (\ref{solha}). Let $s^{\pm},s'^{\pm}$ have $N+1$
`$\pm$' distinct embedding charges and let the sets of these ${\pm}$ embedding charge labels be identical.
Then there exists $|r'\ket$ in $\Psi'$ such that the embedding charge nets 
underlying $r',s$ are identical.
\\

\noindent{\em Proof}: By an appropriate choice of spatial diffeomorphism $\phi$ such that 
$\phi={\bf 1}$ in a small neighbourhood
of   $x=0$ (and hence $x=2\pi$), 
we can arrange for $t= s'_{\phi}$ to be such
that its `+' embedding charge net matches with that of $s$. Consider the coarsest graphs underlying 
$s^-, t^-$. Let their first edges be $e_1,f_1$. 
If $f_1= e_1$, we can proceed to a comparision of the second edges of these charge nets.
If $f_1$ is longer than $e_1$ we can stretch the second edge $f_2$ of 
$t_1$ 
(with charge $k^-_2$) leftwards by repeated applications of {\bf C1} so that on the (coarsest graph underlying the) 
resulting charge net $t^-_1$, 
the first edge is also $e_1$. If $f_1$ is shorter that $e_1$ then we can stretch $f_1$ rightwards by repeated applications
of {\bf N1} so that in the coarsest graph underlying the resulting chargenet $t^-_1$, the first edge is again $e_1$.
\footnote{Note that in applying {\bf C1}, {\bf N1}.  
we are free to 
choose an appropriately fine  graph which underlies $t$.} 

Next, we compare the second edges on $s^-,t^-_1$. If they are unequal, we can use {\bf C1},{\bf N1}
to stretch 
 the 3rd edge of $t^{-}_1$ leftward or the second edge 
rightward so as to generate $t^{-}_2$ on which the first two edges match those
of $s^-$. Clearly iterations of this procedure ensure that $r'^-:=t^{-}_{N-1}$ agrees with $s^-_{N-1}$ while maintaining 
$r'^+= t^+=s^+$. From {\bf C1}, {\bf N1} it follows that 
we have constructed the desired $|r'\ket$ in $\Psi'$.
\\

\noindent {\bf L7}:Let $|s'\ket$ be  a finest lattice charge net in   $\Psi'$ where
$\Psi'$ is a solution 
to the diffeomorphism
constraints and the Hamiltonian constraints (\ref{solh}), (\ref{solha}). Let $s$ be related to $s'$
by a finite gauge transformation $\Phi_+,\Phi_-$. Then $|s\ket$ is also in $\Psi'$.
\\

\noindent{\em Proof}: 
Since $s$ is gauge related to $s'$, it is also a finest lattice state.
The proof of {\bf L4} implies that there exists $|s_1\ket$ which 
is obtained by the action of some ${\hat U}_{\delta, v=0}, {\hat U}_{{\delta}^{\prime}, v=0}$  on $|s\ket$ for 
small enough $\delta, \delta^{\prime}$, 
such that $s_1$  has $N+1$ `+' and $N+1$ `-' embedding charges. From {\bf L2} it follows that if
$|s_1\ket$ is in any solution to the diffeomorphism constraint and to (\ref{solh}),
(\ref{solha}), then $|s\ket$ must be in the same solution.

Next, {\bf L4} implies that there exists $s'_1$ with $N+1$ `+' and $N+1$ `-' embedding charges
such that $|s'_1\ket$ is in $\Psi'$ and 
{\bf L5} implies that there exists $|s'_2\ket$ in $\Psi'$ such that 
$s'_2, s_1$ have identical sets of embedding charges. From {\bf L6}, 
there exists $|s'_3\ket$ in $\Psi'$ such that 
 $s'_3,s_1$ have identical embedding charge networks.

Next, it is straightforward to check that, as asserted in \cite{polypft2}, if two finest lattice states
are gauge related and have the same embedding charge networks, they must have the same matter charge networks.
\footnote{
The interested reader may check that this assertion  follows from property (i) section \ref{sec3} and 
the gauge transformation properties of matter and embedding charge
networks.}
Since the transformations relating $s$ to $s_1$ and $s'$ to $s'_i$, $i=1,2,3$ and $s$ to $s'$ are gauge transformations,
it follows that $s'_3=s_1$. From the first paragraph of this proof it follows that $|s\ket$ is in $\Psi'$.
\\

To summarise: We have shown that if a finest lattice state is in a solution to the diffeomorphism 
and Hamiltonian constraints (\ref{solh}),(\ref{solha}) then all finite gauge transformations of this state
are also in that solution. From section \ref{sec5}, this implies that such a solution encodes long range 
propagation.


\end{document}